\documentclass[reprint,amsmath,amssymb,pra]{revtex4-1}

\usepackage{graphicx}% Include figure files
\usepackage{dcolumn}% Align table columns on decimal point
\usepackage{bm}% bold math
\usepackage[usenames,dvipsnames]{xcolor} 
\usepackage[utf8]{inputenc}
\usepackage{latexsym}
\usepackage{amsfonts}
\usepackage{amssymb}
\usepackage{amsthm}
\usepackage{amsmath}
\usepackage{psfrag}
\usepackage{rotating}
\usepackage{pgf,pgfsys,pgffor} %
\usepackage{pgfplots}
\usepackage{pgfplotstable}
\usepackage{tikz}
\usetikzlibrary{intersections,arrows,decorations.pathmorphing,backgrounds,fit,mindmap,calc,through,scopes,fadings,positioning,automata,calendar,shapes,er,matrix,folding,patterns,petri,plothandlers,plotmarks,shadows,topaths,through,trees,pgfplots.units,decorations, snakes}  
\usepgfplotslibrary{fillbetween}
\usepackage[caption=false]{subfig} % vedi subfigure_vs_subcaption.tex
\usepackage{blkarray}
\newcommand{\trace}[1]{\textrm{Tr} \left[ {#1} \right]}
\newcommand{\bra}[1]{\langle {#1} \vert}
\newcommand{\ket}[1]{\vert {#1} \rangle}
\newcommand{\pure}[1]{\vert {#1} \rangle \langle {#1} \vert}
\newcommand{\inn}[2]{\langle {#1} \vert {#2} \rangle}
\newcommand{\san}[3]{\langle {#1} \vert {{#2}} \vert {#3} \rangle}
\newcommand{\ee}{e} %\newcommand{\ee}{\textrm{e}}			
\newcommand{\dd}{\textrm{d}}	
\newcommand{\ii}{{i\mkern1mu}}				
\newcommand{\vectorize}[1]{\textrm{vec}\left [#1 \right ]} % \vec already exists
\DeclareMathOperator{\sgn}{sgn}

\newcommand{\beq}{\begin{equation}}
\newcommand{\eeq}{\end{equation}}
	
\usepackage[super]{nth}

\usepackage{hyperref}	

\begin{document}

%\preprint{}

\title{Quantum State Discrimination on Reconfigurable Noise--Robust Quantum Networks}
%{Quantum State Discrimination on Reconfigurable Networks via Quantum Stochastic Walks}


%\author[1]{Nicola Dalla Pozza\footnote{Electronic address: nicola.dallapozza@unifi.it (Corresponding author)}}
%\author[1]{Filippo Caruso\footnote{Electronic address: filippo.caruso@unifi.it}}
%\affil[1]{Dept. of Physics and Astronomy \& LENS, University of Florence, Via Carrara 1, I-50019 Sesto Fiorentino, Italy.}

\author{Nicola Dalla Pozza}
\email{nicola.dallapozza@unifi.it}
\author{Filippo Caruso}
\email{filippo.caruso@unifi.it}
\affiliation{Dept. of Physics and Astronomy \& LENS, University of Florence, Via Carrara 1, I-50019 Sesto Fiorentino, Italy.}

\date{\today}

\begin{abstract}
A fundamental problem in Quantum Information Processing is the discrimination amongst a set of quantum states of a system. In this paper, we address this problem on an open quantum system described by a graph, whose evolution is defined by a Quantum Stochastic Walk. In particular, the structure of the graph mimics those of neural networks, with the quantum states to discriminate encoded on input nodes and with the discrimination obtained on the output nodes. We optimize the parameters of the network to obtain the highest probability of correct discrimination. Numerical simulations show that after a transient time the probability of correct decision approaches the theoretical optimal quantum limit. These results are confirmed analytically for small graphs. Finally, we analyze the robustness and reconfigurability of the network for different set of quantum states, and show that this architecture can pave the way to experimental realizations of our protocol as well as novel quantum generalizations of deep learning.
\end{abstract}

%\keywords{}

\maketitle

\section{\label{introduction}Introduction}

In the last decade, Quantum Stochastic Walks (QSW) have been proposed as a model to generalize both quantum walks and classical random walks \cite{Whitfield2010}. Their formulation arises from the need to extend quantum walks to open quantum systems, with the aim of incorporating decoherence effects that are inevitably present in a real physical system. In fact, the formerly proposed quantum version of random walks, i.e. Quantum Walk (QW) \cite{Aharonov1993, Kempe2003, Kendon2007, VenegasAndraca2012, Reitzner2011}, has been defined by a unitary evolution of the state, without taking into account incoherent effects. This allows the walker's position on a graph to be in a superposition of states, a property that has been exploited to show that QWs are universal for quantum computation \cite{Childs2009, Childs2013, Lovett2010} and that they allow to design quantum algorithms with computational advantages over classical algorithms \cite{Ambainis2003, Ambainis2008, Ambainis2010, Giri2017, Santha2008, Farhi1998,Chids2003, Shenvi2003, Childs2004, Ambainis2005}.

In parallel to these results, there have been works showing the beneficial impact of decoherence for dephasing-enhanced transport in a variety of systems, in particular in light-harvesting complexes \cite{Plenio2008, MendozaArenas2013, ContrerasPulido2014, OlayaCastro2008,Caruso2009,Chin2010, Caruso2010}. %Bruderer2016
This has motivated the study of quantum walks with an environmental interaction, as it is described by QSW. The framework of QSW has been investigated in the context of relaxing property \cite{SanchezBurillo2012, Liu2017, Glos2017} and propagation speed \cite{Domino2017, Domino2018}, showing advantages for speed-up in learning algorithms \cite{Schuld2014}, and  enhancement of excitation transport \cite{Mohseni2008, Caruso2014, Viciani2015, Caruso2016, Park2015}.

The evolution of QSW is defined by a Gorini--Kossakowski--Sudarshan--Lindblad master equation \cite{Kossakowski1972, Lindblad1976, Gorini1976},
\beq
\frac{\dd \rho}{\dd t}= -(1-p)\ii \left[ H,\rho \right] + p \sum_{k} \left( L_{k} \rho L_{k}^{\dagger} - \frac{1}{2} \left \{ L_{k}^{\dagger}L_{k}, \rho \right \} \right)
\label{LindbladEq}
\eeq
where one assumes to work in units with $\hbar=1$, $H$ is the Hamiltonian,  $\{L_k\}$ are the Lindblad operators, and both are defined from the adjacency matrix describing the network of nodes involved in the random walk. The smoothing parameter $p$ accounts for the amount of coherent evolution given from $H$ with respect to the irreversible evolution given by the Lindblad operators, and it allows to interpolate between a quantum walk ($p=0$) and a classical random walk ($p=1$).

On the other side, quantum state discrimination has been one of the first problems faced in quantum information theory \cite{Helstrom1976, Holevo1972, Yuen1975, Holevo1979, Chefles2000, Bergou2007, Bergou2010, Barnett2009}, but it is still a flourishing research field as demonstrated from recent theoretical \cite{Sych2016,Croke2017,Rosati2017,Weir2017,Namkung2018, Flatt2019} and experimental works \cite{Muller2012,SolisProsser2017,DiMario2018,Izumi2018,Han2018,Nakahira2018}, also considered in relation to machine learning approaches~\cite{Fanizza2019}. In its most general formulation, an observer wants to guess the quantum state of a system that is prepared in one of a set of feasible states, possibly by optimizing the measurement operators to apply on the system. The performance measure is the 
%. The observer aims at 
%optimizing the coefficients of the measurement operators to solve a discrimination problem, that is, to discern the initial quantum state of the system with the highest 
probability of correct detection 
\beq
P_c = \sum_{n=1}^{\mathcal{M}} p_n \trace{\Pi_n \rho^{(n)}},
\label{def:Pc}
\eeq
where $\{\rho^{(n)}\}$ is the set of quantum states to discriminate, $n=1\ldots \mathcal{M}$,  $\{p_n\}$ their a priori probabilities, and $\{\Pi_n\}$ the measurement operators to estimate them. 

In this work we consider the discrimination of quantum states as a result of their time evolution. Our structure is inspired by neural networks, with its evolution described by a quantum stochastic walk. The connection between the implementation of measurement operators (possibly to solve the discrimination problem) and quantum walks has been investigated in recent works \cite{Kurzynski2013, Li2019}, but in these papers an alternative formulation of quantum walks is used, and we explicitly refer to a neural network structure for the quantum system. 
We test different sets of quantum states and several networks in order to understand the best topologies for the discrimination problem. % We believe we cannot observe the beneficial impact of the noise since the graphs we consider are too small and simple where noise assisted transport is not present since interference effects are neutralized from static disorder in the hamiltonian coefficients (see \cite{Caruso2009, Mohseni2008} and the references therein).

%In the optimized network, we see that the probability of correct decision increases with the total evolution time $\tau$, and that for a fixed $\tau$ the best performances are obtained lowering $p$. An interpretation of these results could be the following: the optimization tries to realize in the network the unitary transformation that rotates the initial quantum states into the measurement basis in the optimal way. \ndp{Decoherence effects can only degrade this process and therefore result in lower performances. Almeno, al momento non sono stati osservati improvements dovuti al rumore}

The paper is organized as follows. In Section \ref{sec:model} we review the formalism of quantum stochastic walks and we introduce the network model that describes the quantum system. We present different topologies for the binary discrimination and for the discrimination between $\mathcal{M}>2$ quantum states. In Section \ref{sec:Discrimination} we formalize the discrimination problem, discussing in Subsection \ref{sec:binaryDiscrimination} the binary case and in Subsection \ref{sec:MaryDiscrimination} the $\mathcal{M}$--ary case. In Section \ref{sec:conclusions} we report the conclusions and final discussions.

\section{Quantum Stochastic walks}
\label{sec:model}

In this section we introduce the QSW model that we here apply to the discrimination problem for the first time.

%A  Quantum Stochastic Walk (QSW) is a generalization of both \emph{classical} random walks and \emph{quantum} walks \cite{Whitfield2010}. 

Classical random walks, quantum walks and quantum stochastic walks are usually defined on an graph $\mathcal{G}$, which is defined by a pair $\mathcal{G}~=~(\mathcal{N},\mathcal{E})$, with $\mathcal{N}$ being a set of elements called nodes (or vertices) and $\mathcal{E}$ being a set of pairs of nodes $(\mathcal{N}_i,\ \mathcal{N}_j)$ representing arcs from $\mathcal{N}_i$ to $\mathcal{N}_j$. The pairs in $\mathcal{E}$ can be summarized in the adjacency matrix $A$, with 
\beq
A_{j, i} = 
\begin{cases} 
1, & \mbox{if } (\mathcal{N}_i,\ \mathcal{N}_j) \in \mathcal{E} \\ 
0, & \mbox{if } (\mathcal{N}_i,\ \mathcal{N}_j) \notin \mathcal{E}
\end{cases} \ .
\label{adjacency}
\eeq
As a generalization, weighted graphs can have any real values $A_{j,i}$ assigned to an arc. Also, in the case the adjacency matrix is a symmetric matrix, i.e., $A_{i,j} = A_{j,i}$, the graph is called \emph{undirected}, otherwise the graph is said to be \emph{directed}. Undirected graphs have pairs $(\mathcal{N}_i,\ \mathcal{N}_j)$ and $(\mathcal{N}_j,\ \mathcal{N}_i)$ with the same weight on them, and in this case the arcs are also called \emph{edges} or \emph{links}.

%are called \emph{directed} if $A_{i,j} \neq A_{i,j}$, and \emph{undirected} otherwise (as f.i. in Eq. \eqref{adjacency}). In a directed graph, we can discriminate the edges $A_{i,j}$ and $A_{j,i} \neq A_{i,j}$ by their ``direction'', with $A_{i,j}$ representing the weight of the link from $N_j$ to $N_i$\footnote{This is true when the probability distribution of the node is represented as a column vector. If the probability distribution is thought as a row vector, $A_{i,j}$ represents the weight of the link from $N_i$ to $N_j$.}. 

The name \emph{random walk} comes from the fact that a walker, starting from an initial node and moving around randomly according to the link connections, assumes a time dependent probability distribution that can be predicted with this framework. In particular, in the case of an undirected graph with equal weights on the edges, we can define the transition--probability matrix $T$ of the possible node transitions as $T=AD^{-1}$, where $D$ is the diagonal degree matrix, with $D_{i,i} = \sum_j A_{j,i}$ representing the number of nodes connected to $i$. The probability distribution of the node occupation, written as a column vector ${\vec q}(t)$, is evaluated for a discrete time~$(t)$ random walk as
\beq
{\vec q}(t+1) = T {\vec q}(t) \ ,
\eeq
and for a continuous time random walk as 
\beq
\frac{\dd {\vec q}}{\dd t} = (T-I) {\vec q} \ .
\eeq

In the quantum scenario, the nodes are associated with the elements of the site basis \cite{Caruso2014} 
% a basis of a quantum system, for instance the site basis 
(see subsection \ref{sec:network} for an extensive description of the network). 
%The probability distribution of the nodes is evaluated with a projective measurement on the corresponding quantum states.
The evolution of the system can be given by the Gorini--Kossakowski--Sudarshan--Lindblad master equation \eqref{LindbladEq},
%\beq
%\frac{\dd \rho}{\dd t} = -(1-p)\ii \left[ H,\rho \right] + p \sum_{i,j}  L_{i,j} \rho L_{i,j}^{\dagger} - \frac{1}{2} \left \{ L_{i,j}^{\dagger}L_{i,j}, \rho \right \} 
%\eeq
with both the Hamiltonian $H$ and the Lindblad operators $\{L_{i,j}\}$ depending on the adjacency matrix defined on the graph. In some models of QSW \cite{Whitfield2010, Caruso2014}, the Hamiltonian operator $H$ is defined from the adjacency matrix, i.e., $H=A$, and with the Lindblad operators depending on the transition matrix defined on the graph as $L_{i,j} = \sqrt{T_{i,j}} \ket{i} \bra{j}$. With this approach, optimizing the coefficients of a weighted undirected adjacency matrix $A$ fixes the Hamiltonian $H$ and the Lindblad operators via $T=AD^{-1}$. Here, we relax this assumption and, once the adjacency matrix defines the topology of the network, 
%that is, to indicate whether a link is present or absent as in Eq. \eqref{adjacency}, but 
we optimize $H$ and $T$ independently. More precisely, the coefficients $A_{i,j}$ are used to decide whether the corresponding $H_{i,j}$ and $T_{i,j}$ will be optimized (independently) or are set to zero. This approach physically corresponds to optimize independently the hopping rates in $H$ and the noise rates in $L$.
Of course, to define proper transition probabilities the matrix $T$ must satisfy a set of constraints, 
\beq
0 \leq T_{i,j} \leq 1 \quad \forall i,j, \qquad \sum_{i} T_{i,j} = 1 \quad \forall j
\label{def:T}
\eeq 
while for simplicity we take $H$ to be any real symmetric matrix, $H_{i,j}=H_{j,i}$, with zero entries on the diagonal. 
%\ndp{Why do we take the assumptions of null diagonal and real (not comlex) coefficients?}

%The parameter $p\in [0,1]$ in \eqref{LindbladEq} acts like a slider to set the amount of unitary dynamics caused by $H$ and the irreversible dynamics diven by $L_{i,j}$. For $p=0$ we obtain a pure quantum walk, while for $p=1$ we obtain a classical random walk. \ndp{già detto, da togliere?}

Moreover, in the graph some nodes may have a particular role. There is usually a starting node that identifies the initial position of the walker. There might be also sink nodes, i.e., nodes that can irreversibly trap the received population. The latter are connected to the rest of the network only through an arc connecting a \emph{sinker} node in the network to the sink, preventing a transition in the reverse direction. In the QSW, this is obtained with a Lindblad operator $L_n = \ket{n}\bra{s_n}$ for each sink, which is added on the right-side of Eq. \eqref{LindbladEq},
\beq
\sum_{n=1}^\mathcal{M} \Gamma \Big (  2 \ket{n}\bra{s_n}\rho \ket{s_n} \bra{n} - \{ \pure{s_n}, \rho \}  \Big )
\label{LindbladSink}
\eeq
%\beq
%\sum_{\hat{k}=1}^m \Gamma \Big (  2 \ket{\hat{k}}\bra{s_{\hat{k}}}\rho \ket{s_{\hat{k}}} \bra{\hat{k}} - \{ \pure{s_{\hat{k}}}, \rho \}  \Big )
%\label{LindbladSink}
%\eeq
with $s_n$ being the sinker node connected to the $n$-th sink $\ket{n}$, $1 \leq n \leq \mathcal{M}$. Overall, the master equation for the density operator $\rho$ describing the system reads
\begin{align}
\frac{\dd \rho}{\dd t} =  & -(1-p)\ii \left[ H,\rho \right] + p \sum_{i,j} L_{i,j} \rho L_{i,j}^{\dagger} - \frac{1}{2} \left \{ L_{i,j}^{\dagger}L_{i,j}, \rho \right \} \notag \\
& \quad + \Gamma \sum_{n=1}^\mathcal{M}   2 \ket{n}\bra{s_n}\rho \ket{s_n} \bra{n} - \{ \pure{s_n}, \rho \},
\label{masterEquation}
\end{align}
and the population at the $n$-th sink at time $t=\tau$ (corresponding to the total evolution time) can be evaluated as
\beq
\rho_{n,n} (\tau) = \bra{n} \rho(\tau) \ket{n} = 2 \Gamma \int_0^\tau \rho_{s_n,s_n}(t)\ \dd t \ .
\label{def:SinkPopulation}
\eeq
From here on, we assume $\Gamma=1$ since this parameter is just a factor defining the time scale ($\Gamma \tau$ is dimensionless). % and can be collected from all the parameters $H_{i,j},\ L_{i,j}$.

\subsection{\label{sec:network}Network model}

To define the topology of the graph we mimic the structure of neural networks \cite{Bishop, Goodfellow, HastieTibshiraniFriedman}. The latter are described by complex graphs where the nodes (resembling neurons) are grouped into input, hidden or output layers. Input nodes are those where the data to be processed are set. Output nodes are those containing the results of the desired task. Hidden nodes represent intermediate steps in the elaboration. %, in some sense ``hidden'' to the user that runs the network. 

%\ndp{li chiamiamo intermediate or hidden?}
% We define a node as being an element of the site basis of the quantum system as in \cite{Caruso2014}. 
In the quantum case a similar network could be physically realized with an ensemble of two--level systems, one for each node, and with a walker realized by a single quantum exciton moving around. Each node is then associated to the state $\ket{i} = \ket{0 \ldots 1 \ldots 0}$, corresponding to have one excitation ($\ket{1}$) at the $i$-th node and $\ket{0}$ elsewhere.
% a product of quantum states $\{ \ket{0},\ \ket{1} \}$ with a $\ket{1}$ at the $i$-th position and $\ket{0}$ elsewhere. 

Then, we classify the nodes of the quantum network into input, intermediate and output layers. For a $M-N-O$ network we mean that there are $M$ input nodes, $N$ intermediate and $O$ output ones (see for instance Fig.\ref{NetworkModels}). Multiple intermediate layers may also be present, for instance a $2-6-5-4$ network has $2$ nodes in the input layer, $6$ nodes in the first intermediate layer, $5$ nodes in the second intermediate one, and $4$ output nodes. The input nodes are associated to a subset of the Hilbert space , where 
%basis, i.e., they define a subspace of the overall quantum system. In this subspace, 
we prepare the initial quantum  (pure or mixed) state of the system, initially in the ground state (no excitons).
% The other nodes are not initialized. Hence, we use the input--nodes subspace to access 
The network will then evolve in time according to Eq. \eqref{masterEquation}. Note that in general the number of input nodes M is not related to the number $\mathcal{M}$ of quantum states to discriminate.
%the corresponding quantum states the initial quantum state is prepared, and the coefficients of all the other quantum states in the basis (corresponding to the intermediate and output nodes) are set to zero.
%The quantum system then evolves according to Eq. \eqref{masterEquation}. 
%Intermediate nodes can be seen as ancillary quantum states involved in the evolution of the quantum system. As already anticipated, the adjacency matrix characterizing the connections between these nodes defines the items of the Hamiltonian and Lindblad operators to be optimized. 
By default, each node in a layer is fully connected with all the nodes of the same layer and with all the nodes of the following one. 
Only in the output layer each sink is connected only from its sinker. 
We also consider different topologies by reducing the connections between nodes within the same layer. 
%Also, by default input layers are not fully connected, but each input node is fully connected with the first intermediate layer. 
When we want to refer to a topology that is not the default one, we use \emph{`r'} to indicate that the connectivity is \emph{reduced}, i.e. some links are removed. The output nodes are sink nodes where the population gets trapped. After the time evolution of the network dynamic, we measure the sink population to estimate the initial quantum state in the discrimination problem (see Section \ref{sec:Discrimination}).

\subsubsection*{Models for binary discrimination}

In the case of binary discrimination, we first consider the  $2r-2r-2$ model (see Fig. \ref{fig:2r-2r-2}). This is probably the simplest model one can imagine, with 2 input nodes connected to 2 sinker nodes and 2 sinks, but with no links between nodes of the same layer. Then, we consider some of its variants obtained adding some links, for instance the $2-2r-2$ model, where the input nodes are connected among themselves, the $2r-2-2$ model, with an additional link between the sinkers, the $2-2-2$ model, with both these links added, and a $2r-4-2$ model, which has 4 intermediate nodes. By comparing the performances of these models we analyze the role of the added edges. In addition, we investigate the role of the intermediate layers optimizing the  $2r-2r-\ldots-2$ model for increasing number of intermediate fully connected layers. This latter model is represented in Fig.~\ref{fig:2r-2r-2r-2}.

\subsubsection*{Models for $\mathcal{M}$-ary discrimination} 

In the case of $\mathcal{M}$-ary discrimination, we consider a setup with the same number of quantum states to discriminate as the input nodes, $\mathcal{M}=M$, and one with a larger number of quantum state, i.e. with $\mathcal{M}>M$. %, with $M$ being the number of input nodes and the size of the subspace spanned by them. 
For instance, we consider the $2-\mathcal{M}-\mathcal{M}$, $2-\mathcal{M}r-\mathcal{M}$ and $2r-\mathcal{M}-\mathcal{M}$ models (see Fig. \ref{fig:2r-4-4}) for $\mathcal{M}=4$ and $\mathcal{M}=8$. As in the binary case, the reduction in the connectivity indicates the absence of links between nodes of the same layer. We also consider $\mathcal{M}r-\mathcal{M}r-\mathcal{M}$, $\mathcal{M}r-\mathcal{M}-\mathcal{M}$ (see Fig. \ref{fig:4r-4-4}) and $\mathcal{M}-\mathcal{M}-\mathcal{M}$ models for $\mathcal{M}=4$.

\tikzstyle{inputNode}=[circle,thick,draw=blue,minimum size=8mm, node distance=15mm]
\tikzstyle{hiddenNode}=[circle,thick,draw=orange,minimum size=8mm, node distance=15mm]
\tikzstyle{outputNode}=[circle,thick,draw=purple,minimum size=8mm, node distance=15mm]
\newcommand{\xshift}{0.8cm}
\begin{figure}[h!]
\subfloat[$2r-2r-2$ model \label{fig:2r-2r-2}]{
	\begin{tikzpicture}[line width=1.2pt]  
		\node[inputNode] (A) {1};
		\node[inputNode, below of=A] (B) {2};
		\node[hiddenNode, right of=A, xshift=\xshift] (C) {3}
			edge (A) edge (B);
		\node[hiddenNode, below of=C] (D) {4}
			edge (A) edge (B);
		\node[outputNode, right of=C, xshift=\xshift] (E) {5}
			edge [pre] (C);
		\node[outputNode, below of=E] (F) {6}
			edge [pre] (D);
		\node[above of=A, anchor=east] {INPUT};
		\node[above of=C] {INTERMEDIATE};
		\node[above of=E, anchor=west, xshift=-3mm] {OUTPUT (SINK)};
	\end{tikzpicture} 
}%
\\[2mm]
\subfloat[$2r-2r-2r-2$ model\label{fig:2r-2r-2r-2}]{
	\begin{tikzpicture}[line width=1.2pt]  
		\node[inputNode] (A) {1};
		\node[inputNode, below of=A] (B) {2};
		\node[hiddenNode, right of=A, xshift=\xshift] (C) {3}
			edge (A) edge (B);
		\node[hiddenNode, below of=C] (D) {4}
			edge (A) edge (B);
		\node[hiddenNode, right of=C, xshift=\xshift] (E) {5}
			edge (C) edge (D);
		\node[hiddenNode, below of=E] (F) {6}
			edge (C) edge (D);
		\node[outputNode, right of=E, xshift=\xshift] (G) {7}
			edge [pre] (E);
		\node[outputNode, below of=G] (H) {8}
			edge [pre] (F);
	\end{tikzpicture}
}%
\\[2mm]
\subfloat[$2r-4-4$ model\label{fig:2r-4-4}]{
	\begin{tikzpicture}[line width=1.2pt]  
		\node[inputNode] (A) {1};
		\node[inputNode, below of=A] (B) {2};
		\node[hiddenNode, right of=A, xshift=\xshift] (D) {4}
			edge (A)	edge (B);
		\node[hiddenNode, above of=D] (C) {3}
			edge (A) 	edge (B) edge (D);			
		\node[hiddenNode, right of=B, xshift=\xshift] (E) {5}
			edge (A)	edge (B) edge [bend right=45] (C) edge (D);
		\node[hiddenNode, below of=E] (F) {6}
			edge (A)	edge (B) edge [bend right=45] (C) edge [bend right=45] (D) edge (E);
		\node[outputNode, right of=C, xshift=\xshift] (G) {7}
			edge [pre] (C);
		\node[outputNode, right of=D, xshift=\xshift] (H) {8}
			edge [pre] (D);
		\node[outputNode, right of=E, xshift=\xshift] (I) {9}
			edge [pre] (E);
		\node[outputNode, right of=F, xshift=1cm] (L) {10}
			edge [pre] (F);						
	\end{tikzpicture} 
}%
\\[2mm]
	\subfloat[$4r-4-4$ model\label{fig:4r-4-4}]{
		\begin{tikzpicture}[line width=1.2pt]  
			\node[inputNode] (A) {1};
			\node[inputNode, below of=A] (B) {2};
			\node[inputNode, below of=B] (C) {3};
			\node[inputNode, below of=C] (D) {4};
			\node[hiddenNode, right of=A, xshift=\xshift] (E) {5}
				edge (A)	edge (B) edge (C) edge (D);
			\node[hiddenNode, right of=B, xshift=\xshift] (F) {6}
				edge (A)	edge (B) edge (C) edge (D) edge (E);
			\node[hiddenNode, right of=C, xshift=\xshift] (G) {7}
				edge (A)	edge (B) edge (C) edge (D) edge (F) edge [bend right=45] (E);
			\node[hiddenNode, right of=D, xshift=\xshift] (H) {8}
				edge (A)	edge (B) edge (C) edge (D) edge (G) edge [bend right=45] (E) edge [bend right=45] (F);
			\node[outputNode, right of=E, xshift=\xshift] (I) {9}
				edge [pre] (E);
			\node[outputNode, right of=F, xshift=\xshift] (L) {10}
				edge [pre] (F);
			\node[outputNode, right of=G, xshift=\xshift] (M) {11}
				edge [pre] (G);
			\node[outputNode, right of=H, xshift=\xshift] (N) {12}
				edge [pre] (H);					
		\end{tikzpicture} 
	}
\caption{Examples of $M-N-O$ network models. In each subfigure, the left layer (blue) collects the input nodes, the right layer (purple) the output nodes while in between (orange) the intermediate nodes, which can be organized in multiple layers as in Fig.\ref{fig:2r-2r-2r-2}. Directed edges refer to irreversible transfer of population, while plain links indicate both coherent and incoherent transport. By default the nodes of a layer are fully connected within themselves. We indicate with \emph{r} when the connectivity is \emph{reduced}, i.e. some links are removed.
%specify irreversible-only dynamics, plain lines indicate both coherent and irreversible evolution. 
}
\label{NetworkModels}
\end{figure}
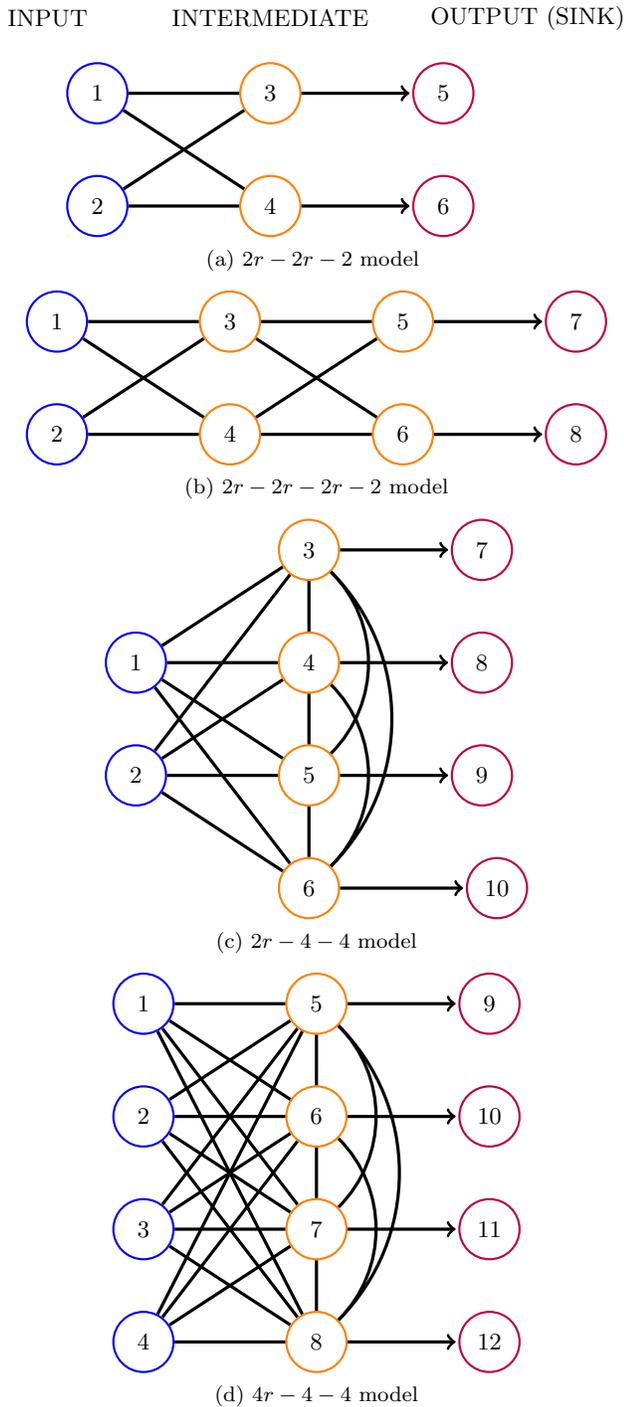

\section{\label{sec:Discrimination} Quantum State Discrimination}

In this section we introduce the problem of quantum state discrimination (for more details see 
%. For a deeper insight into the topic, we suggest the reader to refer to the 
reviews \cite{Chefles2000, Bergou2007, Bergou2010, Barnett2009}).

Assume that a quantum system is prepared in a quantum state drawn from a set of given states, represented by density operators $\{\rho^{(m)}, 1\leq m \leq \mathcal{M}\}$ in the Hilbert Space $\mathcal{H}$. The a--priori probabilities $p_m$ by which the quantum states are prepared are also known, i.e. $\{p_m, 1\leq m \leq \mathcal{M}\},\ \sum_m p_m=1$. In the discrimination problem, we search for the Positive Operator-Valued Measurements operators (POVM) $\{\Pi_n, 1\leq n \leq \mathcal{M}\}$ satisfying
\beq
\Pi_n  \geq 0, \quad \sum_{n=1}^\mathcal{M} \Pi_n = I_\mathcal{H},
\label{POVMconstraints}
\eeq
that allow to estimate the prepared state with the highest probability of correct detection $P_c$, or equivalently, the lowest probability of error $P_e=1-P_c$. In \eqref{POVMconstraints} the term $I_\mathcal{H}$ represents the identity operator acting on the space of the density matrices of the quantum system $\mathcal{H}$.

If the quantum states $\{\rho^{(m)}\}$ span orthogonal subspaces, a perfect discrimination is possible by appropriate measurement operators and $P_c=1$. If this is not the case, the outcome corresponding to the measurement operator $\Pi_n$ may be correctly recorded when the prepared quantum state is $\rho^{(n)}$, or it may be wrongly recorded when the quantum state is $\rho^{(m)}$, $m~\neq~n$, leading to the 
probability of correct detection in Eq.~\eqref{def:Pc}.
% With the quantum state $\rho^{(m)}$, the outcome probability of the estimate $n$ is $\trace{\Pi_n \rho^{(m)}}$, and therefore the probability of correct detection is given by
%\beq
%P_c = \sum_{n=1}^\mathcal{M} p_n \trace{\Pi_n \rho^{(n)}}
%\label{def:Pc}
%\eeq

The conditions for the optimal solution have been derived by Holevo \cite{Holevo1972} and by Yuen, Kennedy and Lax \cite{Yuen1975}. In the most general scenario the problem can be solved numerically via semidefinite programming \cite{Eldar2003}, but, in the binary case or if the set of quantum states exhibits symmetry features, the optimization can be further carried on analytically to better understand the structure of the measurement operators \cite{Eldar2004, Nakahira2013, DallaPozza2015}, and possibly to find a closed form for the probability of correct decision.

In the case of only two quantum states, the problem of binary discrimination has been solved by Helstrom \cite{Helstrom1976}, and the optimal probability of correct decision is known as \emph{Helstrom bound}. If the quantum states are pure, $\rho^{(1)} = \pure{\psi^{(1)}}$ and $\rho^{(2)} = \pure{\psi^{(2)}}$, the Helstrom bound reads
\beq
P_c^{Helstrom} = \frac{1}{2} \left (1+\sqrt{1-4p_1p_2 \vert\inn{\psi^{(1)}}{\psi^{(2)}}\vert^2}\right).
\label{PcHelstrom}
\eeq
If instead $\rho^{(1)}$ and $\rho^{(2)}$ are mixed, the Helstrom bound can be evaluated numerically \cite{Helstrom1976}. When the discrimination is set among $\mathcal{M}>2$ quantum states, the theoretical optimal probability of correct decision ${P_c}^*$ is evaluated numerically, and it is used as a reference for the performance of the network. 
%\beq
%P_c^{Helstrom} = \frac{1}{2} \left( 1+\sqrt{1-\cos(2\theta)^2 }\right) = \frac{1+\sin(2\theta)}{2}
%\label{PcOptimal}
%\eeq

Note that in our setup we actually consider an equivalent formulation of the problem, where instead of optimizing the measurement operators we fix the measurement projectors (on the population of the output nodes) and optimize the evolution of the network. 
%The measurement is described by the projectors associated with the quantum states of the output nodes, i.e., with their representation in the site basis. 
The optimization of the evolution of the system concerns the coefficients of the Hamiltonian and the Lindblad operators to obtain the best evolution from the subspace of the input nodes, where the quantum states to discriminate are prepared, to the subspace of the output nodes, where the measurement is performed. The two problems are equivalent, and one can interpret the optimized evolution with the measurement on the output nodes as realizing the Naimark extension of the POVM defined on the input nodes for the original discrimination problem.
%Note that in our setup, we actually consider an equivalent formulation of the problem, where instead of optimizing the measurement operators we fix them to be the projectors defined by the site basis corresponding with the output nodes of the network, and we optimize the \emph{``transformation''} from the subspace of the input nodes to the subspace of the output nodes. The problem is equivalent, and we can interpret this transformation as realizing the Naimark extension of the POVM for the original discrimination problem. 
To be more precise, to have a resolution of the identity as in Eq. \eqref{POVMconstraints} we formally need to include a projection on the subspace outside the output nodes. This is necessary since part of the population can be trapped in the network \cite{Caruso2009, Caruso2014}. The outcome associated to this extra projector is considered inconclusive for the discrimination.
%We could in principle ignore the projections on input and intermediate nodes if we could assure the complete evolution of the initial quantum states onto the subspace of the output nodes. As we will see, we cannot make such assumption for a finite time evolution, and we have to consider projections on all the nodes, with the possibility to get outcomes corresponding to nodes that are not sinks. In this latter case we will consider the outcome inconclusive.

%\ndp{ottimalità dei proiettori (kennedy) se gli stati da discriminare sono linearmente indipendenti
%---> non se ne discrimino M che spannano 2.
%assumendo che tutta la popolazione vada nei sink , è come fare una discriminazione sullo spazio dei sink con una rotazione data dall'evoluzione temporale. 
%anche se non riesco a realizzare i POVM, magari i proiettori corrispondono a quelli dell'esteinsione di naimark.}

We fix the measurement operators to be $\{\Pi_n = \pure{n}\}$, with $n$ identifying the $n$-th sink,
%. Within these projectors, those associated with the $n$-th sink are renamed $\Pi_i = \Pi_n$ with 
$1\leq n \leq \mathcal{M}$, which is associated with the estimation of the input quantum state $\rho^{(n)}$. The projector associated with the inconclusive output is $\Pi_{inc} = I_\mathcal{H} - \sum_n \Pi_n$. % The other projectors $\Pi_i,\ i\neq n,$ give inconclusive results \footnote{We can actually aggregate all these projectors associated with inconclusive results into one.}.

The probability of correct decision can then be written as
\begin{align}
P_c & = \sum_{n=1}^\mathcal{M} p_n\ \trace{\Pi_n \rho^{(n)} (\tau)} = \sum_{n=1}^\mathcal{M} p_n\ \rho_{n,n}^{(n)}(\tau) \notag \\
%& = 2\Gamma \sum_{n=1}^\mathcal{M} p_n  \int_0^\tau \trace{\Pi_{s_n} \rho^{(n)}(t)} \ \dd t \notag \\
& = 2\Gamma \sum_{n=1}^\mathcal{M} p_n  \int_0^\tau \rho_{s_n,s_n}^{(n)}(t) \ \dd t 
\label{def:PcNetwork}
\end{align}
%\begin{align}
%P_c & = \sum_{n=1}^\mathcal{M} \trace{\Pi_n \rho_n p_n} = \sum_{n=1}^\mathcal{M} \rho_{n,n\vert n} p_n \notag \\
%& = \sum_{n=1}^\mathcal{M} p_n  \int_0^\tau \trace{\Pi_{s_n} \rho^{(n)}(t)} \dd t \notag \\
%& = \sum_{n=1}^\mathcal{M} p_n  \int_0^\tau \rho_{s_n,s_n \vert n}(t) \dd t \notag \\
%\end{align}
where we have defined $\rho_{i,j}^{(m)}(t)$ as
\beq
\rho_{i,j}^{(m)}(t) = \bra{i}\rho^{(m)}(t) \ket{j}.
\eeq
The optimization of the network coefficients has been performed numerically using
standard routines % \cite{BoydVandenberghe} .
employing an interior--point algorithm \cite{BoydVandenberghe, Byrd1999,Byrd2000,Waltz2006} to maximize the probability of correct decision $P_c$.
%Following the practice used in machine learning to solve classification problems \cite{Bishop, Goodfellow, HastieTibshiraniFriedman}, we consider as cost function the average cross-entropy error between the unknown quantum states $m$ and the estimate $n$ rather than the probability of correct decision,
%\beq
%CrossEntropy(m,n)=-\sum_m \sum_n - \delta_{n,m} p_m \log \left \{\trace{\Pi_n \rho^{(m)} (\tau)} \right \} = -\sum_n p_n \log[\rho_{n,n}^{(n)}(\tau)],
%\label{def:crossEntropy}
%\eeq
%obtained after evaluating the time evolutions of all the initial quantum states and the corresponding probability of correct detection $\trace{\Pi_n \rho^{(n)} (\tau)}=\rho_{n,n}^{(n)}(\tau)$. 
%At each iteration of the optimization algorithm, for each quantum state to discriminate $\rho^{(n)}(0)$ the final quantum state $\rho^{(n)}(\tau\)$ is evaluated, and a measurement is performed to evaluate the corresponding probability of correct detection $\trace{\Pi_n \rho^{(n)} (\tau)}=\rho_{n,n}^{(n)}(\tau)$. 
%Once the network has been optimized, we evaluate its performance with the probability of correct decision $P_c$.
%\ndp{REF for relation between crossentropy ed error probability?}

In the case of the simple network $2r-2r-2$ for the binary discrimination, we further carry on the optimization analytically and solve the problem for $p=0$ and $p=1$. This has given us some insights on how to interpret the behaviour of the performance as a function of $p$ and $\tau$. More details are reported in the Appendices \ref{vectorization}--\ref{sec:solutionP1}.

%If you use MSE, the weight adjustment factor (the gradient) contains a term of (output) * (1 – output). As the computed output gets closer and closer to either 0.0 or 1.0 the value of (output) * (1 – output) gets smaller and smaller. For example, if output = 0.6 then (output) * (1 – output) = 0.24 but if output is 0.95 then (output) * (1 – output) = 0.0475. As the adjustment factor gets smaller and smaller, the change in weights gets smaller and smaller and training can stall out, so to speak.
%
%But if you use cross-entropy error, the (output) * (1 – output) term goes away (the math is very cool). So, the weight changes don’t get smaller and smaller and so training isn’t s likely to stall out. Note that this argument assumes you’re doing neural network classification with softmax output node activation.
%

%\section{ Numerical solution for $0<p<1$}
%
%condizioni di ottimalità che si vedono..
%confronto ottimizzazione numerica con quella con le condizioni di ottimalità

\subsection{\label{sec:binaryDiscrimination}Binary discrimination}

%In this section we show the results of the network optimization for the binary discrimination. 
We set up the discrimination problem with different pairs of states, for different values of $p \in [0,1]$ and of the total evolution time $\tau$ in Eqs~\eqref{masterEquation}, \eqref{def:SinkPopulation}. For each pair  $(p,\tau)$ we optimize $H$, $T$ in Eqs~\eqref{def:T}, \eqref{masterEquation} assuming equal a--priori preparation probabilities of the states to get discriminated.

\begin{figure*}[p]
\subfloat[$\protect\begin{pmatrix}
			0.8536 & 0.3536 \\
			0.3536 & 0.1464
			\protect\end{pmatrix}$ vs $\protect\begin{pmatrix}
			0.8536 & -0.3536 \\
			-0.3536 & 0.1464
			\protect\end{pmatrix}$ \label{fig:pureStatesNoSy}]{%
	\centering
	\includegraphics[width=0.9\columnwidth]{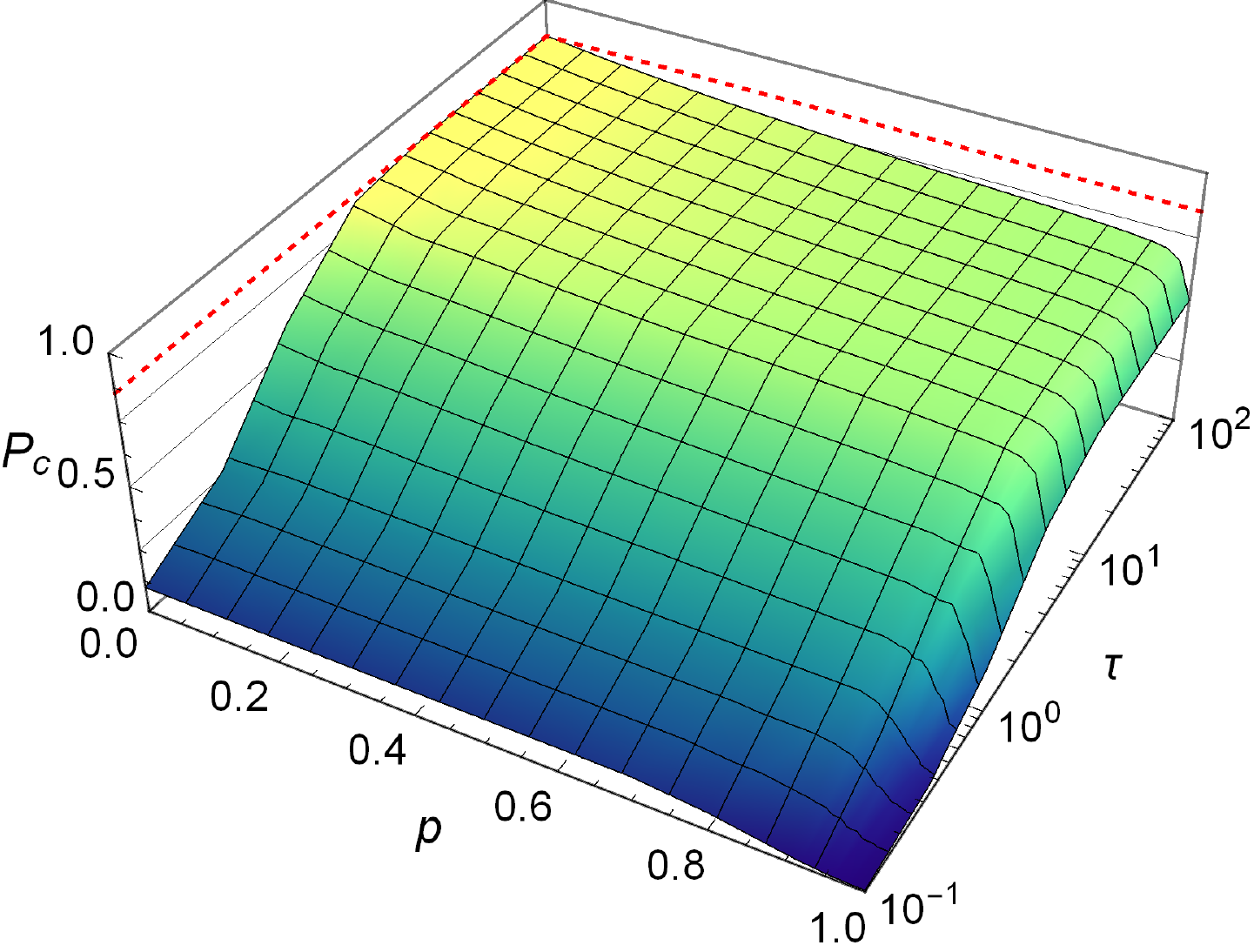}
}~
\subfloat[$\protect\begin{pmatrix}
	0.8536 & - \ii 0.3536 \\
	\ii 0.3536 & 0.1464
	\protect\end{pmatrix}$ vs $\protect\begin{pmatrix}
	0.8536 & i0.3536 \\
	- i0.3536 & 0.1464
	\protect\end{pmatrix}$ \label{fig:pureStatesWithSy}]{%
	\centering
	\includegraphics[width=0.9\columnwidth]{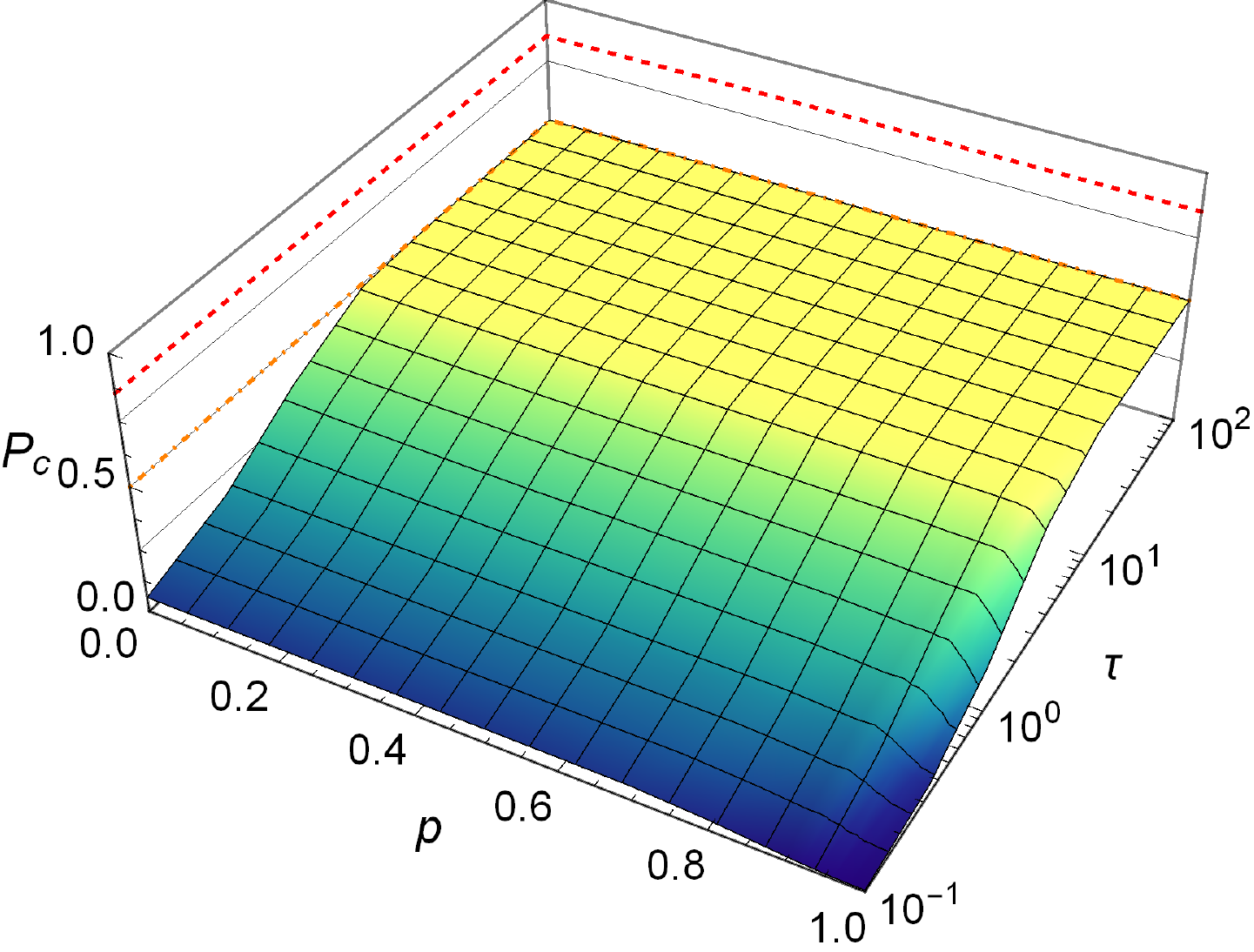}
}%
\\[2mm]
\subfloat[$\protect\begin{pmatrix}
	0.8536 & -0.3536 \\
	-0.3536 & 0.1464
	\protect\end{pmatrix}$ vs $\protect\begin{pmatrix}
	0.6768 & 0.1768 \\
	0.1768 & 0.3232
	\protect\end{pmatrix}$ \label{fig:pureVsMixedNoSy}]{%
	\centering
	\includegraphics[width=0.9\columnwidth]{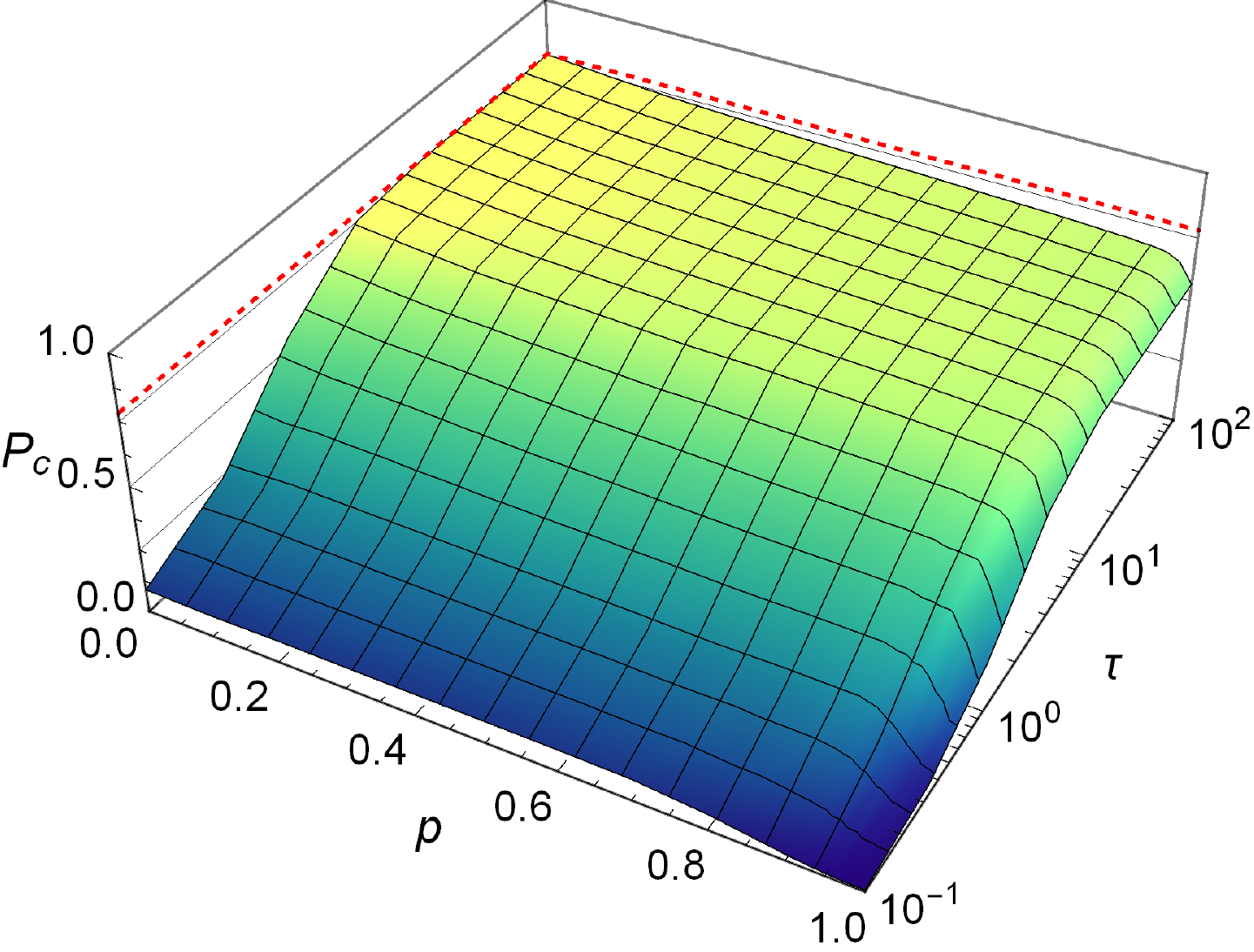}
}~
\subfloat[$\protect\begin{pmatrix}
	0.8536 & i0.3536 \\
	-i0.3536 & 0.1464
	\protect\end{pmatrix}$ vs $\protect\begin{pmatrix}
	0.6768 & -i0.1768 \\
	i0.1768 & 0.3232
	\protect\end{pmatrix}$ \label{fig:pureVsMixedWithSy}]{%
	\centering
	\includegraphics[width=0.9\columnwidth]{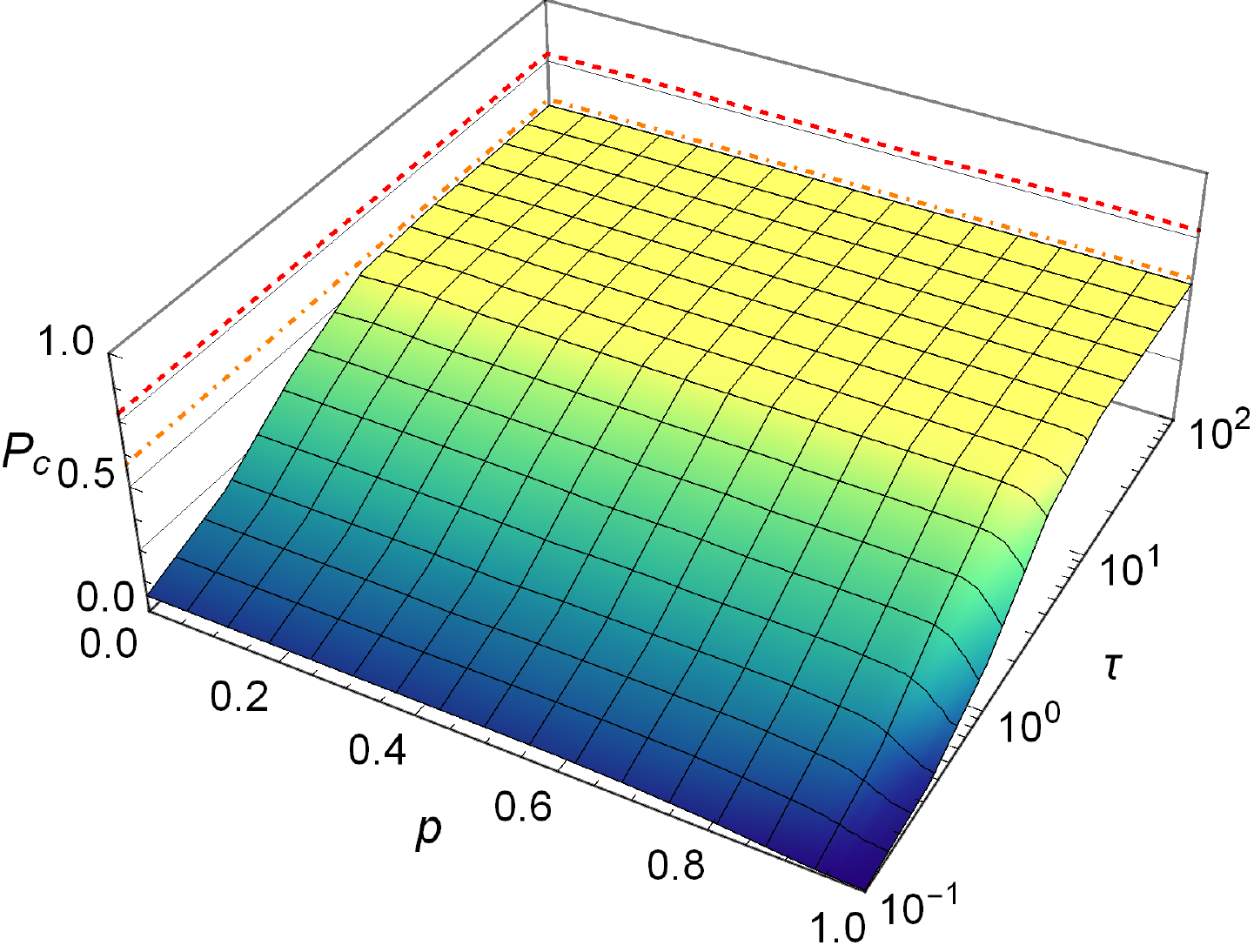}
}%
\\[2mm]
\subfloat[$\protect\begin{pmatrix}
	0.6768 & - 0.1768 \\
	-0.1768 & 0.3232
	\protect\end{pmatrix}$ vs $\protect\begin{pmatrix}
	0.6768 & 0.1768 \\
	0.1768 & 0.3232
	\protect\end{pmatrix}$ \label{fig:mixedStatesNoSy}]{%
	\centering
	\includegraphics[width=0.9\columnwidth]{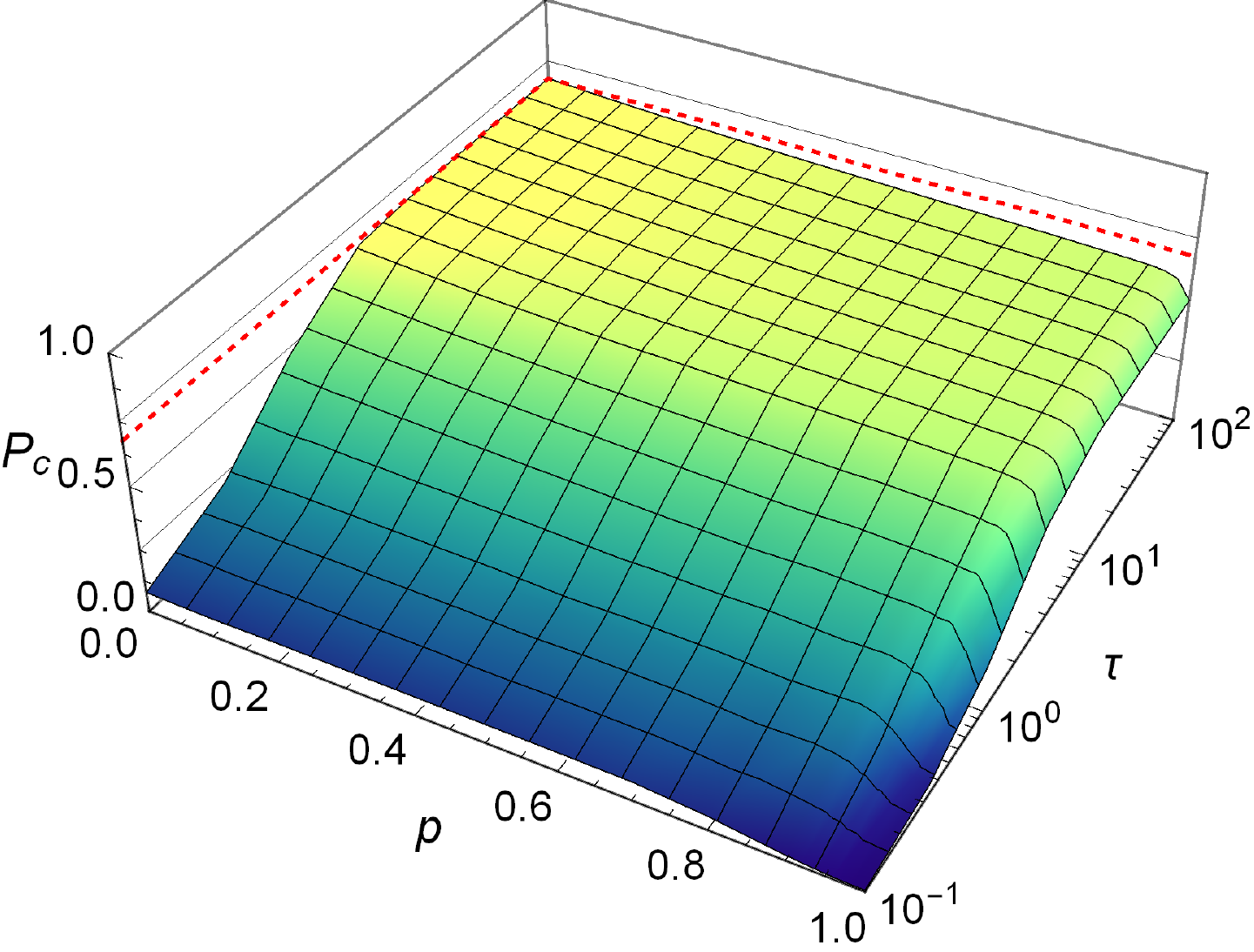}
}~
\subfloat[$\protect\begin{pmatrix}
	0.6768 & - i0.1768 \\
	i0.1768 & 0.3232
	\protect\end{pmatrix}$ vs $\protect\begin{pmatrix}
	0.6768 & i0.1768 \\
	- i0.1768 & 0.3232
	\protect\end{pmatrix}$ \label{fig:mixedStatesWithSy}]{%
	\centering
	\includegraphics[width=0.9\columnwidth]{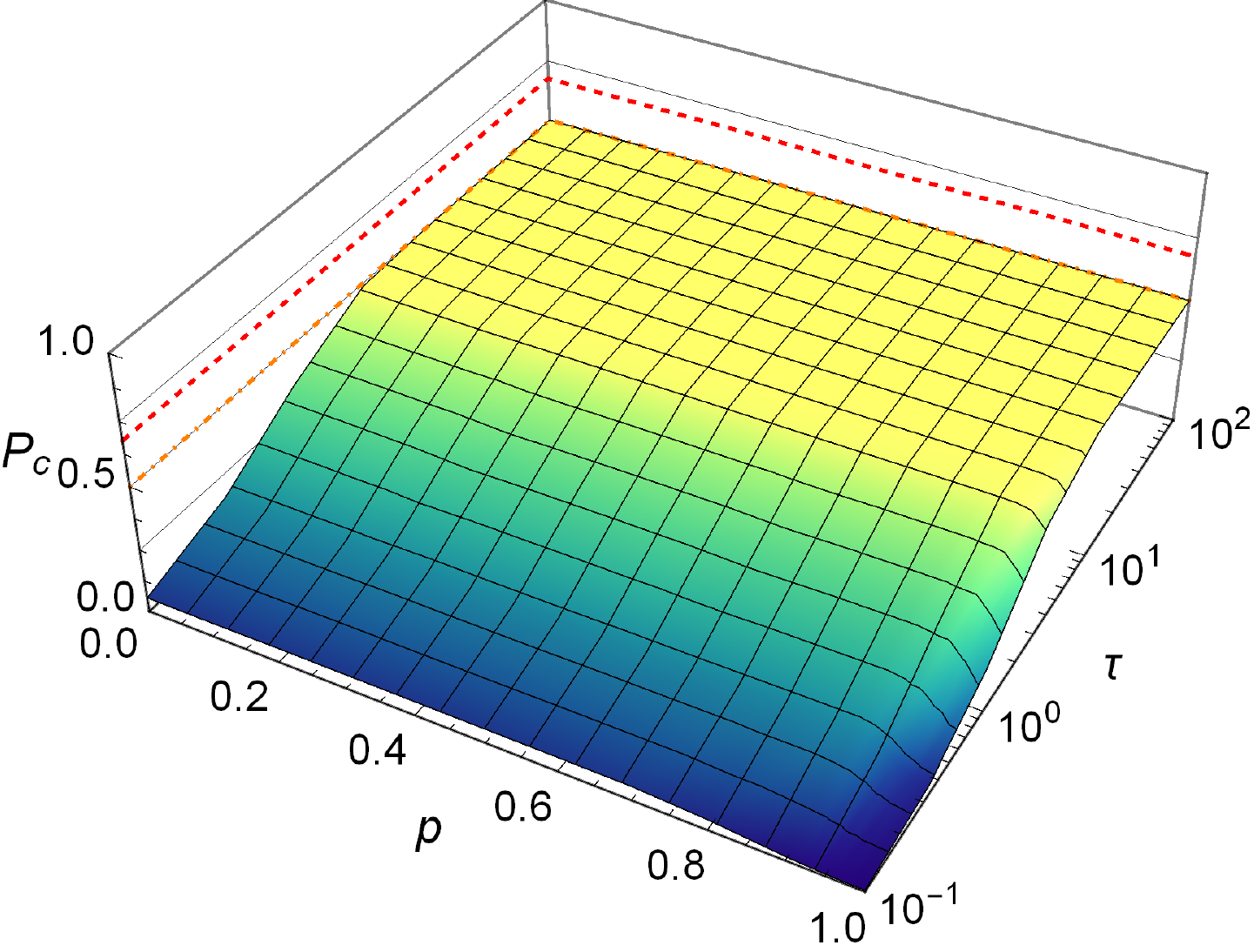}
}%
\caption{Optimized probability of correct decision with respect to $(p,\ \tau)$ for the $2r-2r-2$ model in the case of binary discrimination with pure states (\nth{1} row), a pure state and a mixed state (\nth{2} row) and mixed states (\nth{3} row). The quantum states, reported in each subcaption, on the left column (a-c-e) have $r_x \neq 0,\ r_y=0$, while those in the right column (b-d-f) have been rotated to have the $r_y$ coordinates in place of the $r_x$ ones. The Helstrom bound is reported in red dashed line, and in dash-dotted orange line we evaluate the Helstrom bound on the same quantum states after setting $r_y=0$. The color map is linear between blue and yellow and normalized between the minimum and the maximum point in each plot. 
}
\label{fig:models_2r-2r-2}
\end{figure*}

Firstly, we consider a $2r-2r-2$ model. We discriminate between two pure states that are symmetric with respect to $\ket{1}$, 
\begin{align}
\ket{\psi^{(1)}} &= \cos \theta \ket{1} + \sin \theta \ket{2}, \notag\\
\ket{\psi^{(2)}} &= \cos \theta \ket{1} - \sin \theta \ket{2}, 
\label{def:pureStates}
\end{align}
and 
\begin{align}
\ket{\psi^{(1')}} &= \cos \theta \ket{1} + \ii \sin \theta \ket{2}, \notag\\
\ket{\psi^{(2')}} &= \cos \theta \ket{1} - \ii \sin \theta \ket{2}, 
\end{align}
with $\theta = \pi/8$.
The probability of correct decision $P_c$ is shown as a function of $(p,\tau)$ in Figs.~\ref{fig:pureStatesNoSy},~\ref{fig:pureStatesWithSy}, with $\rho^{(1)}=\pure{\psi^{(1)}}$, $\rho^{(2)}=\pure{\psi^{(2)}}$ reported below each plot.
We also consider the discrimination between the pure state $\ket{\psi^{(1)}}\ \big [\ket{\psi^{(1')}} \big]$ and a mixed state $\rho^{(2)}\ \big [\rho^{(2')}\big ]$ with the same spherical coordinates $(r_x,\ r_y,\ r_z)$ in the Bloch sphere representation $\rho^{(2)}=(1+r_x \sigma_x + r_y \sigma_y + r_z \sigma_z)/2$ but with a radius $r=\sqrt{r_x^2+r_y^2+r_z^2}$ reduced to 0.5 (this value has been chosen to be intermediate between that of a pure state and the completely mixed state). The plots are shown in Figs. \ref{fig:pureVsMixedNoSy}, \ref{fig:pureVsMixedWithSy}.
Then, we consider the discrimination between the mixed quantum states $\rho^{(1)},\ \rho^{(2)} \ \big [\rho^{(1')},\ \rho^{(2')}\big ]$ obtained from $\ket{\psi^{(1)}},\ \ket{\psi^{(2)}}\ \big [\ket{\psi^{(1')}},\ \ket{\psi^{(2')}}\big ] $ reducing both radii to 0.5. The plots are shown in Figs. \ref{fig:mixedStatesNoSy}, \ref{fig:mixedStatesWithSy}.
Note that the quantum states $\ket{\psi^{(1')}},\ \ket{\psi^{(2')}},\ \rho^{(1')},\ \rho^{(2')}$ are simply obtained by rotating $\ket{\psi^{(1)}},\ \ket{\psi^{(2)}},\ \rho^{(1)},\ \rho^{(2)}$ in the Bloch Sphere in order to change the $r_x$ coordinates into the $r_y$ coordinates. We find that for increasing values of $\tau$ the performance increases. This can be interpreted by the fact that an initial quantum state requires some time to reach the sinks. In addition, we can see an almost-constant negative slope in $p$ for a fixed $\tau$, with the quantum walk ($p=0$) outperforming the general quantum stochastic walk with $p>0$. We can also notice that the performance seems to saturate asymptotically, for any $p$, approaching the Helstrom bound for $p=0$ in the plot (a), (c) and (e). The plots (b), (d), (f) also show a clear gap between the surface and the optimal theoretical bound.

%. Also, while in the left plots the asymptotic limit corresponds to the Helstrom bound for $p=0$, on the right plots we can see a clear gap between the surface and the optimal theoretical bound.

We have further investigated this behaviour solving the optimization problem analytically for $p=0$ and $p=1$. We provide the expression for the optimal $H,\ T$, in the Appendices \ref{vectorization}, \ref{invariant} and \ref{sec:solutionP0}, and we prove that for $p=0$ asymptotically we can reach the Helstrom bound while for $p=1$ we reach the theoretical classical bound, that is, the Helstrom bound evaluated on the quantum states with the coherences set to zero.

Now we give a sketch of the solution in the particular case $p=0$ and the discrimination between the pure quantum states \eqref{def:pureStates}, $\theta \in [0, \pi/2]$, on a $2r-2r-2$ model. Assuming the Hamiltonian in the form
\beq
H=\begin{pmatrix}
0 & 0 & h & h \\
0 & 0 & h & -h \\
h & h & 0 & 0 \\
h & -h & 0 & 0
\end{pmatrix},
\eeq
we can evaluate the time evolution of the node population (more details in Appendix \ref{sec:solutionP0}) and the probability of correct decision results
\begin{align}
& P_c(\tau) = \int_0^\tau \rho_{3,3}^{(1)}(t) + \rho_{4,4}^{(2)}(t)\ \dd t \notag \\
& =  \frac{1+\sin(2\theta)}{2} \left[1-e^{-\tau} \left(\frac{z \sinh \left(z \tau\right)+\cosh \left(z \tau\right)-1}{z^2}+1\right)\right]
\end{align}
with $z=\sqrt{1-8 h^2}$. The maximization of $P_c(\tau)$ for a finite $\tau$ requires the minimization of the term $f(z) = (z \sinh \left(z \tau\right)+\cosh \left(z \tau\right)-1)/z^2$, which can be accomplished numerically. In the asymptotic limit of $\tau \to \infty$ the term in the brackets vanishes, and
$P_c(\infty)$ equals the Helstrom bound evaluated on $\ket{\phi_1},\ \ket{\phi_2}$. 

While in the case of the quantum states \eqref{def:pureStates} the optimized network approaches the optimal performance for $p=0$ in the asymptotic limit, this is not the case for any pair of quantum states. In fact, the $2r-2r-2$ model has an invariant subspace \cite{Caruso2009} not connected to any sink. An invariant subspace of a quantum system dynamics is an Hilbert subspace where the dynamics is confined, i.e., span of the eigenstates of $\mathcal{H}$ that are orthogonal to the output nodes. In the case of the $2r-2r-2$ model, the invariant subspace is present for any $p$ and includes the $y$-component of the state in the Bloch sphere, that is, the $r_y$ component of the quantum state starts in this invariant subspace and its evolution remains trapped there. This means that the problem is equivalent to discern the quantum states after setting $r_y$ to zero, effectively projecting the quantum states in the $(\sigma_x, \ \sigma_z)$ plane of the Bloch sphere (see Fig.~\ref{fig:models_2r-2r-2} right panel). 

%The theoretical optimal performances (Helstrom bound) for these quantum states are also reported in Fig.~\ref{fig:models_2r-2r-2} in orange dash-dotted line.
%Defining the quantum states without these $\sigma_y$ coherences, we would obtain the theoretical optimal performances reported with the orange dash-dotted line in the plots on the right column, which are approached asymptotically.

\begin{figure}[h!]
	\centering
	\includegraphics[width=0.9\columnwidth]{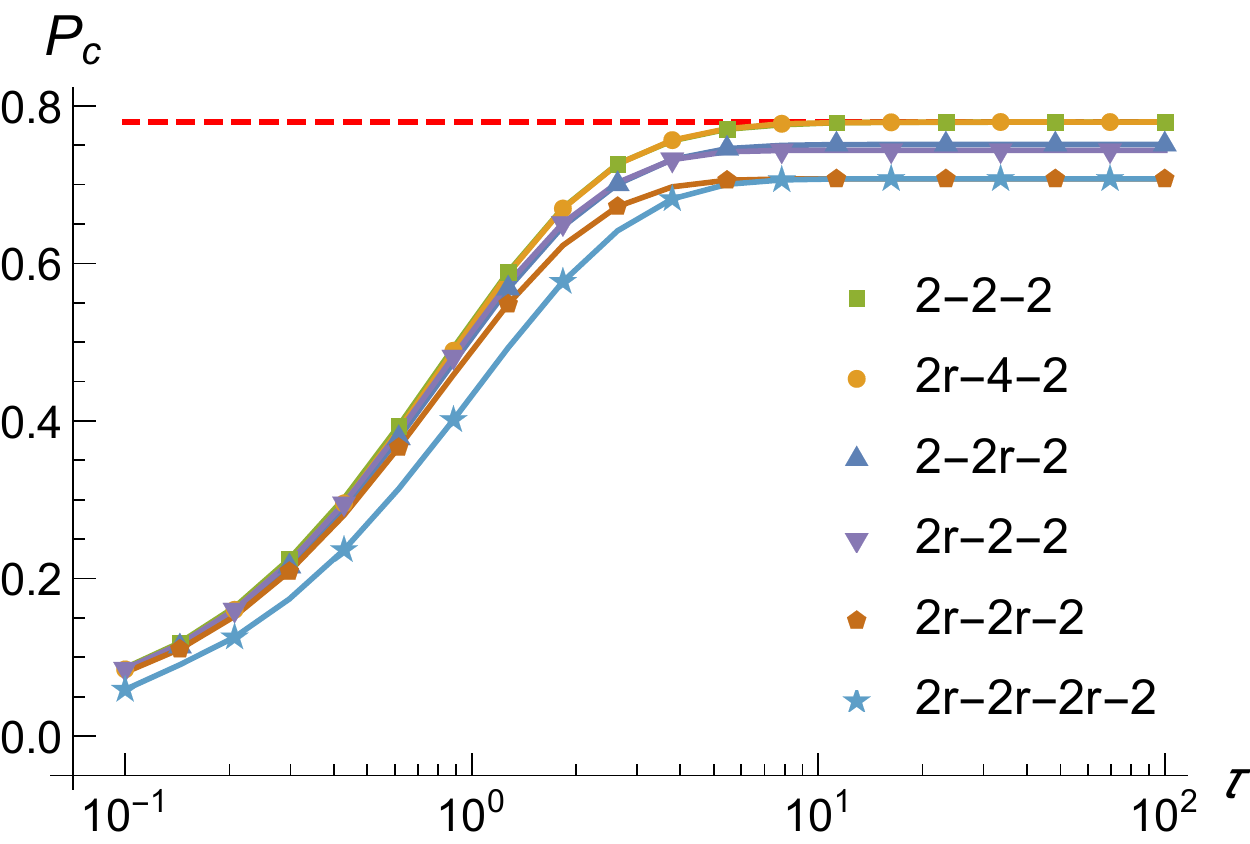}
	\caption{Probability of correct detection $P_c$ for different variants of the  $2r-2r-2$ model as a function of $\tau$ for $p=0$. 
	%The quantum states to discriminate are reported in Eq.~\eqref{binary_QS_simulations}
	%$\rho^{(1)}~=~\protect\begin{pmatrix}
	%0.8536 & 0.25 - \ii 0.25 \\
	%0.25 + \ii 0.25 & 0.1464
	%\protect\end{pmatrix}$ 
	%and 
	%$\rho^{(2)}~=~\protect\begin{pmatrix}
	%0.6768 & -0.125 - \ii 0.125 \\
	%-0.125 + \ii 0.125 & 0.3232
	%\protect\end{pmatrix}$. 
	The red dashed line shows the Helstrom bound for mixed quantum states in Eq.~\eqref{binary_QS_simulations}
	}
\label{fig_binary_topologies}
\end{figure}

Secondly, we investigate the role of some links in the network performance, by evaluating the performance of some variants of the $2r-2r-2$ model. For instance, we add a link in the input layer ($2-2r-2$ model), a link in the intermediate layer ($2r-2-2$ model), a link in both layers ($2-2-2$ model), an intermediate layer ($2r-2r-2r-2$ model) and some intermediate nodes in the same layer ($2r-4-2$ model). We compare the performances in Fig.~\ref{fig_binary_topologies}, where we optimize the discrimination between a pure state and a mixed state, both with some $r_y$ coordinates, 
\begin{align}
%\rho^{(1)} = \pure{\psi^{(1)}}=\begin{pmatrix}
	%0.8536 & 0.25+\ii 0.25 \\
	%0.25-\ii 0.25 & 0.15
	%\protect\end{pmatrix}, \quad 
%=\begin{pmatrix}
	%0.6768 & -0.125 - 0.125i \\
	%-0.125 + 0.125 i & 0.3232
	%\protect\end{pmatrix}	
\rho^{(1)} & = \pure{\psi^{(1)}},\ \ket{\psi^{(1)}} = \cos \theta \ket{1} + \sin \theta \ee^{-\ii \xi}\ket{2} , \label{binary_QS_simulations} \\
\rho^{(2)} & = \frac{1 - r \cos \xi \sin 2\theta \sigma_x + r \sin \xi \sin 2\theta \sigma_y + r \cos 2\theta \sigma_z}{2}, \notag 
%\rho^{(2)} & = \frac{1 + r \cos \xi \sin 2\theta \sigma_x + r \sin \xi \sin 2\theta \sigma_y + r \cos 2\theta \sigma_z}{2}, \notag 
\end{align}
with $\theta=\pi/8,\ \xi = \pi/4,\ r=0.5$. These values have been chosen to have both $r_x$ and $r_y$ coordinates, with an intermediate radius between 1 (corresponding to a pure state) and 0 (corresponding to the completely mixed state). The probability of correct decision as a function of $(p,\tau)$ has the same general behaviour of Fig. \ref{fig:models_2r-2r-2}. For this reason, in Fig. \ref{fig_binary_topologies} we plot $P_c$ for $p=0$ as a function of $\tau$ for the different models, along with the Helstrom bound. As in the previous comparison, the performance increase in $\tau$ and the saturation threshold can be clearly observed. Indeed, it is interesting to compare the saturation value amongst the models. The models $2r-2r-2$ and $2r-2r-2r-2$ have similar performances, showing a gap with the Helstrom bound in the asymptotic value. This is due to the presence of invariant subspaces trapping the quantum state component corresponding to the $r_y$ coordinates. Interestingly, the addition of a link in the input or intermediate layer of models $2r-2-2$ and $2-2r-2$, despite breaking the invariant subspaces, allow for an increase of the performance but do not close the gap with the Helstrom bound. Finally, the models $2-2-2$ and $2r-4-2$ approach the upper bound. This is particularly interesting because it suggests that a reduced topology in the input nodes could be compensated by an increased number of intermediate nodes in a single layer.

We believe we cannot observe the beneficial impact of the noise since the graphs we consider are too small and simple where noise assisted transport is not present since interference effects are neutralized from static disorder in the Hamiltonian coefficients (see \cite{Caruso2009, Mohseni2008} and the references therein).

%\begin{figure}[h!]
%\subcaptionbox{Model $2r-2-2$ \label{fig:2r-2-2}}{%
	%\centering
	%\includegraphics[width=0.45\columnwidth]{Pc_2r-2-2_mesh.eps}
%}%
%\hfill
%\subcaptionbox{Model $2-2r-2$ \label{fig:2-2r-2}}{%
	%\centering
	%\includegraphics[width=0.45\columnwidth]{Pc_2-2r-2_mesh.eps}
%}%
%\\[2mm]
%\subcaptionbox{Model $2r-2-2-2$ \label{fig:2r-2-2-2}}{%
	%\centering
	%\includegraphics[width=0.45\columnwidth]{Pc_2r-2-2-2_mesh.eps}
%}%
%\hfill
%\subcaptionbox{Model $2r-4-2$ \label{fig:2r-4-2}}{%
	%\centering
	%\includegraphics[width=0.45\columnwidth]{Pc_2r-4-2_mesh.eps}
%}%
%\caption{Probability of correct detection for different variations of the model $2-2-2$. The red dashed line shows the Helstrom bound.
%}
%\label{fig:binaryPM}
%\end{figure}

\subsubsection{Robustness}

Here, we analyze the robustness of our discrimination scheme in the case of noisy preparation of the quantum states, a noisy configuration of the network and also an increasing number of intermediate layers.

\begin{figure}[h!]
\centering
	\includegraphics[width=0.9\columnwidth]{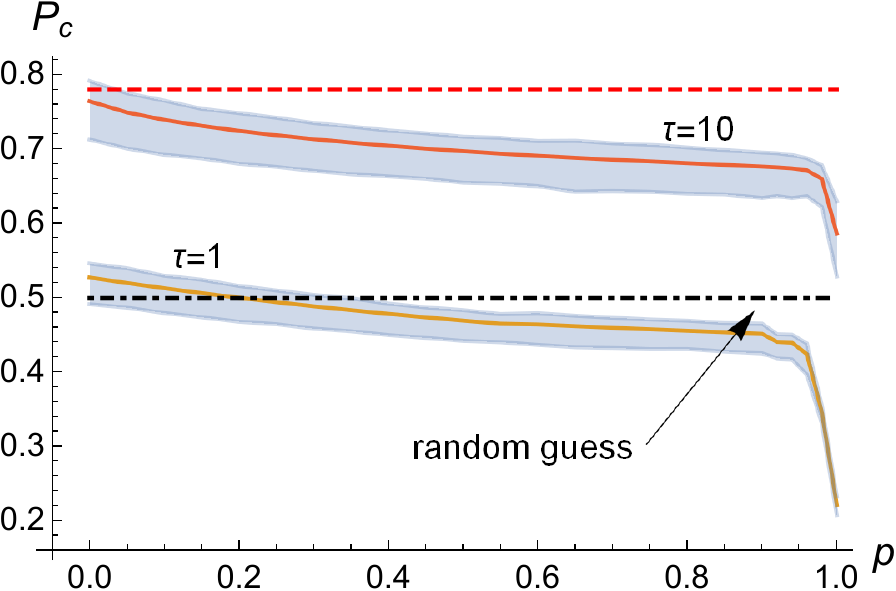}
\caption{Probability of correct decision $P_c$ (blue region) around the predicted performance (solid lines) with a noisy preparation of the quantum states to discriminate ($5\%$ of percent error). The red dashed line shows the Helstrom bound. For each $p$ and $\tau=1,\ 10$ we run $10^4$ evaluations of $P_c$, each with a different pair of quantum states.
}
\label{fig_robustness_states}
\end{figure}

\begin{figure}[h!]
\centering
	\includegraphics[width=0.9\columnwidth]{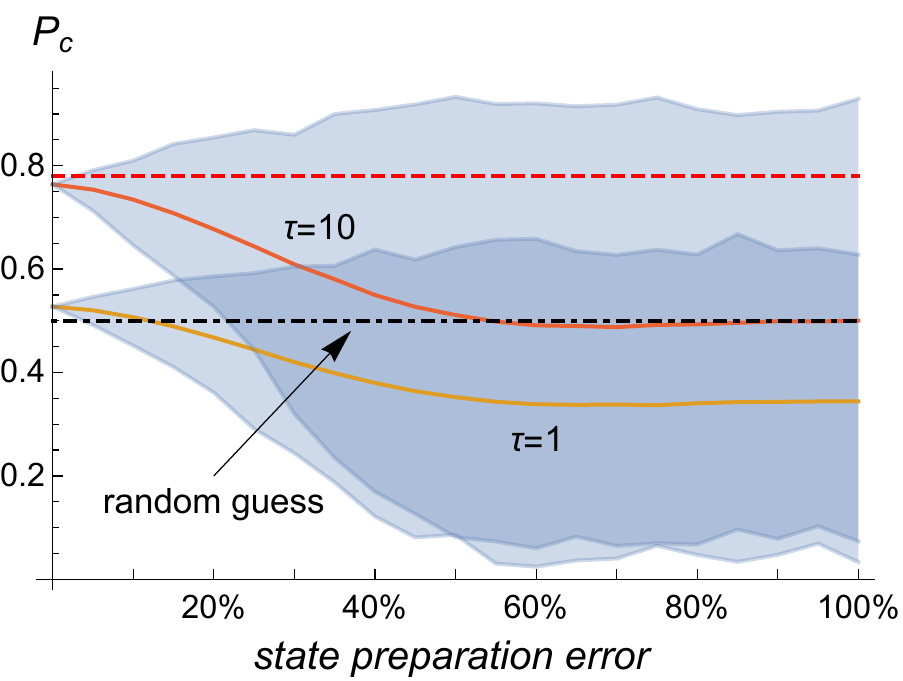}
\caption{Probability of correct decision $P_c$ (blue region) as a function of the maximum percent error on the preparation of quantum states, for $p=0$ and $\tau=1,\ 10$. The continuous line shows the average performance over $10^4$ simulations. The black dash--dotted line shows the threshold corresponding to a random guess (0.5). The red dashed line shows the Helstrom bound.
}
\label{fig_robustness_states_percentError}
\end{figure}

In the former case, we optimize the $2-2-2$ model assuming to discriminate the quantum states in \eqref{binary_QS_simulations} while only a noisy preparation is actually available, for instance due to experimental imperfections in the preparation stage. We prepare the network with the optimal setup, i.e., with the optimal coefficients in $H,\ L$, but we input two random quantum states $\rho^{(1)},\ \rho^{(2)}$ by uniformly sampling $\theta,\ \xi,\ r$ around their nominal value with a maximum of $5\%$ percent error. We run $10^4$ simulations with this setup as a function of $p$ for $\tau=1,\ 10$, sampling new pairs of quantum states at each run. In Fig. \ref{fig_robustness_states} we show that even with a noisy preparation of the quantum states, for any $p$ the performances remain close to the theoretical values. Additionally, we focus on how the performance varies as a function of the preparation error ranging from $0\%$ to $100\%$ for $p=0$ -- see  Fig.~\ref{fig_robustness_states_percentError}. We find that up to around $5\%$ of error the correct decision probability remains very close to the Helstrom bound and anyway decreases linearly with the preparation error up to around $25\%$, before exponentially dropping down below the random guess case (i.e. $P_c = 0.5$).
%Note that although on average the performance worsen, the sampled quantum states could be more distinguishable and exhibit better probability of correct decision. For reference we plot a line at $P_c=0.5$ which indicates the threshold of the random guess. 


\begin{figure}[h!]
\subfloat[$p=0$ \label{fig_robustness_HTcoefficients_percentError}]{
\centering
	\includegraphics[width=0.9\columnwidth]{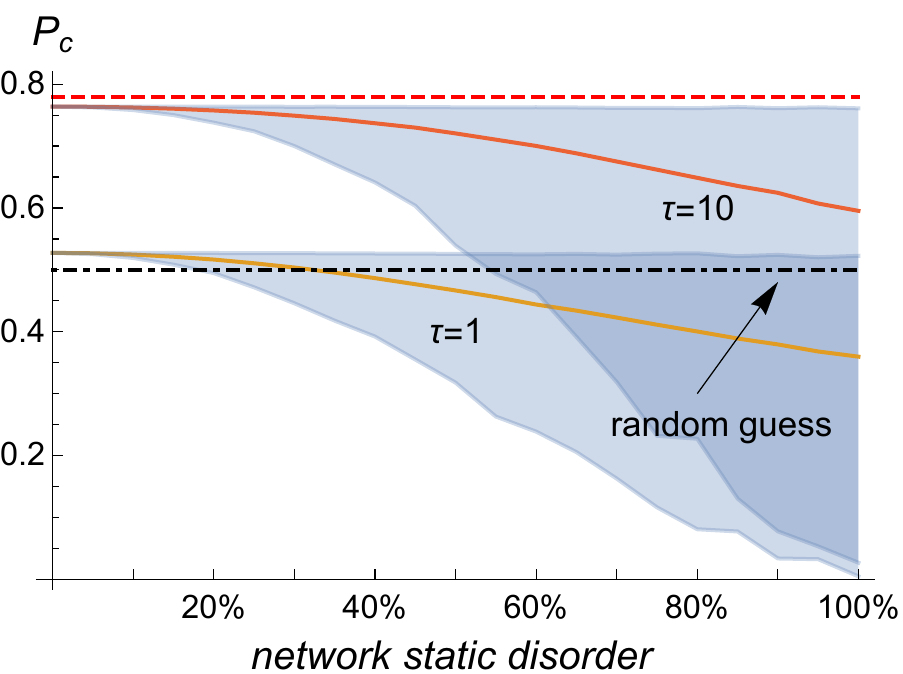}
}%
\\[2mm]
\subfloat[$p=0.1$ \label{fig_robustness_HTcoefficients_percentError_p01}]{
\centering
	\includegraphics[width=0.9\columnwidth]{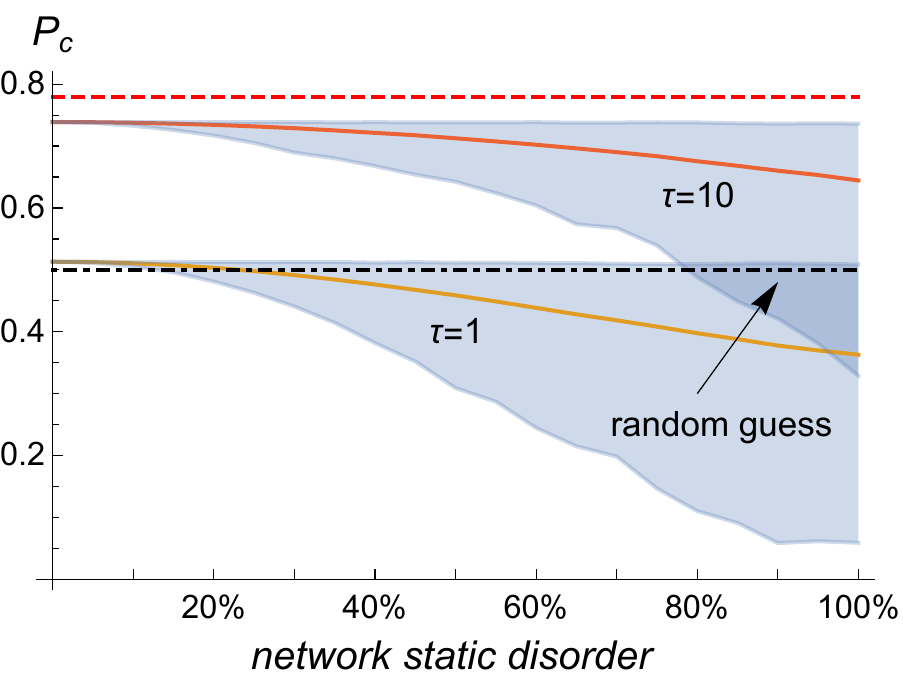}
}%
\caption{Probability of correct decision $P_c$ (blue region) for a network prepared in a noisy configuration of its Hamiltonian coefficients $H_{i,j}$, as a function of its network static disorder percentage, for $p=0$ (a), $p=0.1$ (b) with $\tau=1,\ \tau=10$. The solid lines show the average performance over $10^4$ runs. The black dash--dotted line shows the threshold corresponding to a random guess (0.5). The red dashed line shows the Helstrom bound.
}
\label{fig_robustness_HTcoefficients}
\end{figure}

As a second robustness test, we consider a perfect preparation of the quantum states of Eq.~\eqref{binary_QS_simulations} but a noisy setup of the network coefficients. In this case, we focus on $p=0,\ 0.1$ and $\tau=1,\ 10$, and in each simulation we sample the coefficients of the Hamiltonian $H$ uniformly around the optimal values with a given maximum percent error corresponding to a sort of network static disorder. Figure \ref{fig_robustness_HTcoefficients} shows that indeed the discriminatory network is robust against noisy preparations of the network dynamics, due for instance to experimental imperfections. It is also interesting to compare the two plots corresponding to the cases with $p=0$ and $p=0.1$. The former case achieves asymptotically a better performance with small network static disorder. The probability of correct decision slowly decreases as a function of this error and only at around $50\%$ of error it approaches the random guess limit. However, the $p=0.1$ case has a more robust performance with respect to this disorder, with the performance range crossing the threshold of random guess at around $80\%$. Towards an experimental implementation of our protocol, it could be beneficial to consider $p=0.1$, slightly sacrificing the performances in favor of a more robust discrimination.
%We run $10^4$ simulations and plot the results in Fig.~\ref{fig_robustness_HTcoefficients_percentError} as a function of the maximum percent error ranging from $0\%$ to $100\%$. The solid lines represent the average performance in the simulations, while the blue regions surrounding them shows the corresponding performance range. For reference, we plot the Hestrom bound and the threshold corresponding to a random guess.

\begin{figure}[h!]
\centering
	\includegraphics[width=0.9\columnwidth]{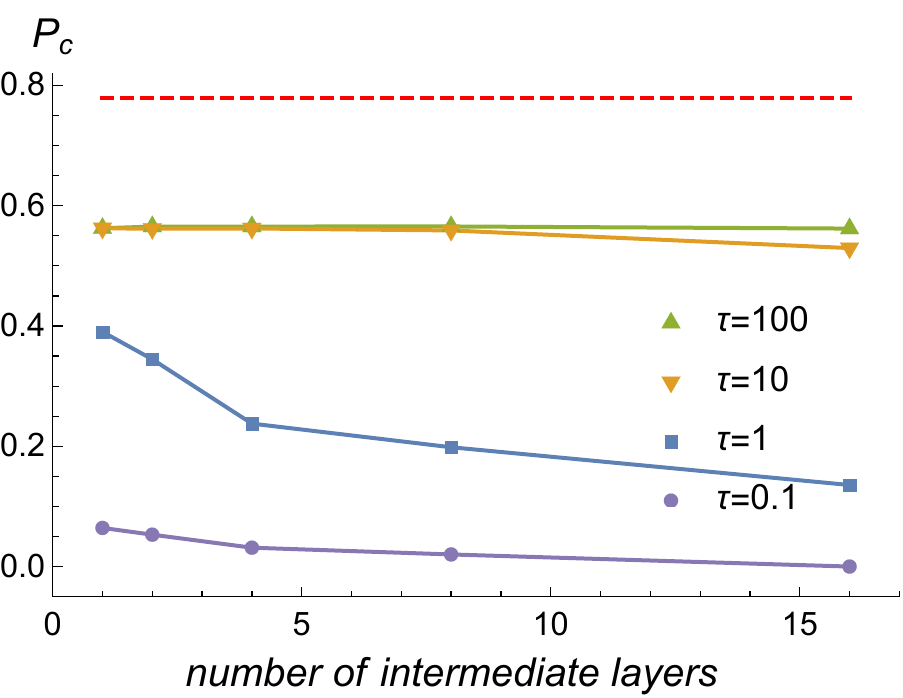}
\caption{Probability of correct decision between quantum states of Fig. \ref{fig:pureVsMixedWithSy} in the model $2r-2r-\ldots-2$ as a function of the  number of intermediate layers, for $\tau \in \{0.1, 1, 10, 100\}$. The (red) dashed line shows the Helstrom bound.
}
\label{fig_robustness_layers}
\end{figure}

As a third robustness test we study whether the probability of correct decision improves adding or removing more intermediate layers. We consider $p=0$ and the $2r-2r-2$ model with 1, 2, 4, 8 and 16 intermediate layers, and we plot the performance for different evolution times in Fig. \ref{fig_robustness_layers}. It turns out that the increased amount of layers does not allow to close the gap with the Helstrom bound.  As the number of intermediate layers increases, the performance lowers due to the fact that a deeper network, i.e., a network with more layers, requires more time to move the quantum state from the input nodes to the output ones. Comparing Figs \ref{fig_binary_topologies} and \ref{fig_robustness_layers}, we find that to increase the discrimination performance it is more convenient to add links inside an intermediate layer rather than increasing the depth of the network.

\subsection{\label{sec:MaryDiscrimination}$\mathcal{M}$-ary discrimination}

We now consider the generalization of our scheme to the discrimination of $\mathcal{M}$ quantum states. In particular, we investigate whether the number of nodes $M$ in the input layer poses limitations in the distinguishability of the quantum states. In general, $M$ and $\mathcal{M}$ are not related, meaning that it could be $M<\mathcal{M}$, $M=\mathcal{M}$ or $M>\mathcal{M}$.
%Regarding the $\mathcal{M}$-ary discrimination, we consider first a set of $\mathcal{M}$ quantum states defined on an input layer of size $M$, i.e., with $M$ nodes. In particular, 

As a first case, we consider the discrimination of the $\mathcal{M}$ pure qubits ($M=2$) 
\beq
\ket{\psi^{(m)}} = \frac{\ket{1} + \ee^{\ii 2 \pi \frac{m}{\mathcal{M}}}\ket{2}}{\sqrt{2}} , \quad m=1 , \ldots , \mathcal{M}
\label{def:equiPhaseStates}
\eeq
with $\mathcal{M}=4, \ 8 $. In the Bloch sphere representation, these quantum states are equally spaced along the equator defined by the $\trace{\rho \sigma_z}~=~0$ plane, and because of this symmetry they are often used to test discrimination protocols \cite{Helstrom1976, DallaPozza2019}.

\begin{figure}[h!]
\subfloat[$\mathcal{M}=4$ \label{fig:2-4-4}]{%
	\centering
	\includegraphics[width=0.9\linewidth]{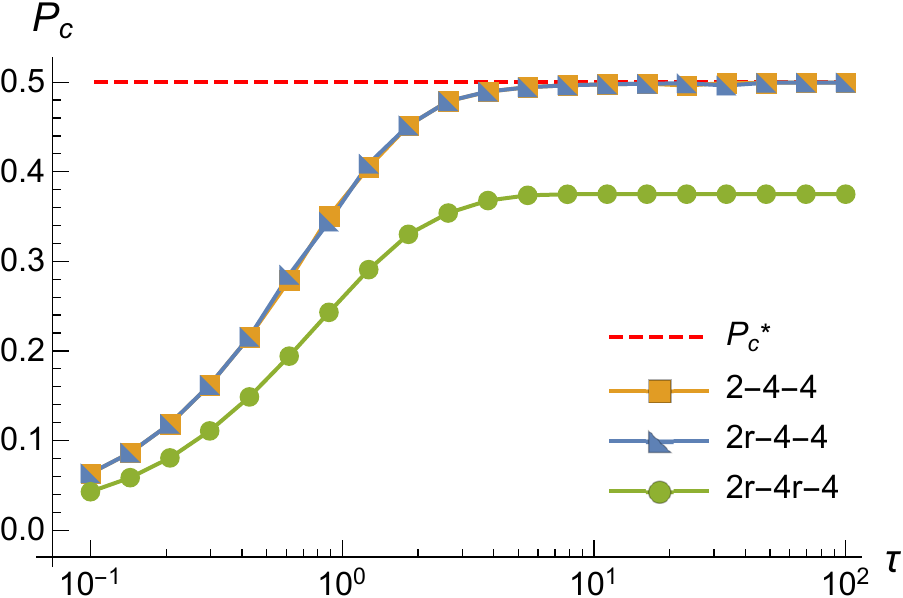}
}%
\\[2mm]
\subfloat[$\mathcal{M}=8$ \label{fig:2-8-8}]{%
	\centering
	\includegraphics[width=0.9\linewidth]{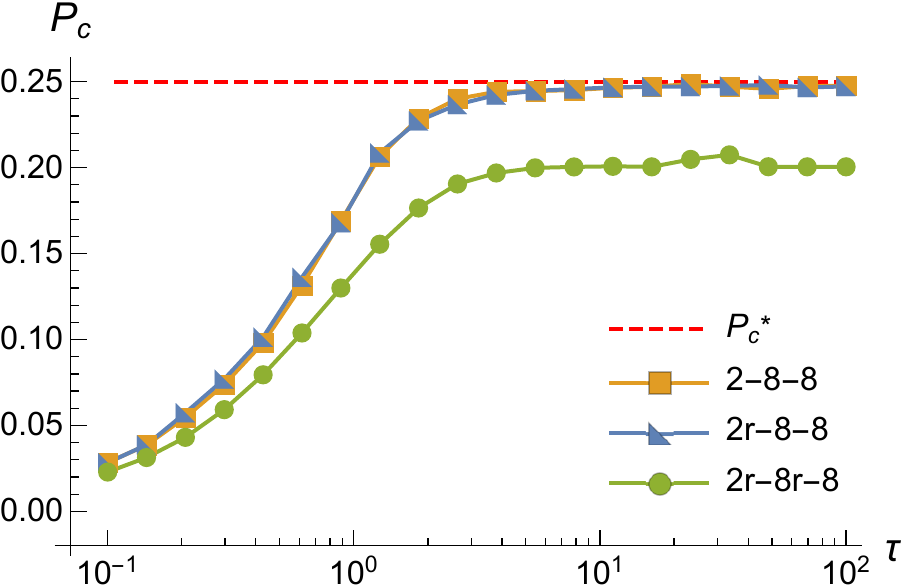}
}
\caption{Probability of correct detection $P_c$ of pure qubits \eqref{def:equiPhaseStates} for $\mathcal{M}=4$ (a) and $\mathcal{M}=8$ (b) as a function of time $\tau$ for different variants of a $2-\mathcal{M}-\mathcal{M}$ model. 
}
\label{fig:models_2-4-4}
\end{figure}

We consider the models $2r-\mathcal{M}r-\mathcal{M}$, $2r-\mathcal{M}-\mathcal{M}$ and $2-\mathcal{M}-\mathcal{M}$, whose performance are reported in Fig. \ref{fig:models_2-4-4}, where we plot only the behaviour for $p=0$ since the trend with respect to $p$ and $\tau$ is similar to the binary case. In this figure
%. \ref{fig:models_2-4-4} 
we just focus on understanding whether the topology asymptotically closes the gap with the optimal bound ${P_c}^*=1-2/\mathcal{M}$, which has been reported for instance in \cite{Helstrom1976}.

There is however a fundamental difference here with respect to the binary case. While in the latter the optimal measurement operators are projectors, here the optimal ones are given by POVM. We can realize these POVM via projectors in an \emph{extended} Hilbert space using the Naimark theorem \cite{Naimark1943, Helstrom1976, DallaPozza2017, DallaPozza2019}, meaning that the optimal network will try to implement such extended projectors via its dynamics and the measurement on the sink nodes. We find that $2-\mathcal{M}-\mathcal{M}$ is the most general model and has the highest performance, approaching asymptotically the optimal ${P_c}^*$. Interestingly, the model $2r-\mathcal{M}-\mathcal{M}$ share the same behaviour, while for $2r-\mathcal{M}r-\mathcal{M}$ the performances are clearly lower.
%
%, and such extension is called and the optimal network tries to implement such projector 
%approaching 
 %with the projectors given by the sink nodes \ndp{REF naimark}. , at least asymptotically for $p=0$, the unitary evolution to realize 

%Even though our measurement on the output nodes it performed with projectors, the optimal network This means that asymptotically, for $p=0$ the optimal network is implementing the unitary evolution to realize the POVM Naimark extension with the projectors corresponding to the sink nodes \ndp{REF naimark}. 

\begin{figure}[h!]
\subfloat[$\alpha = 0.3$ \label{fig:4-_03}]{%
	\centering
	\includegraphics[width=0.9\linewidth]{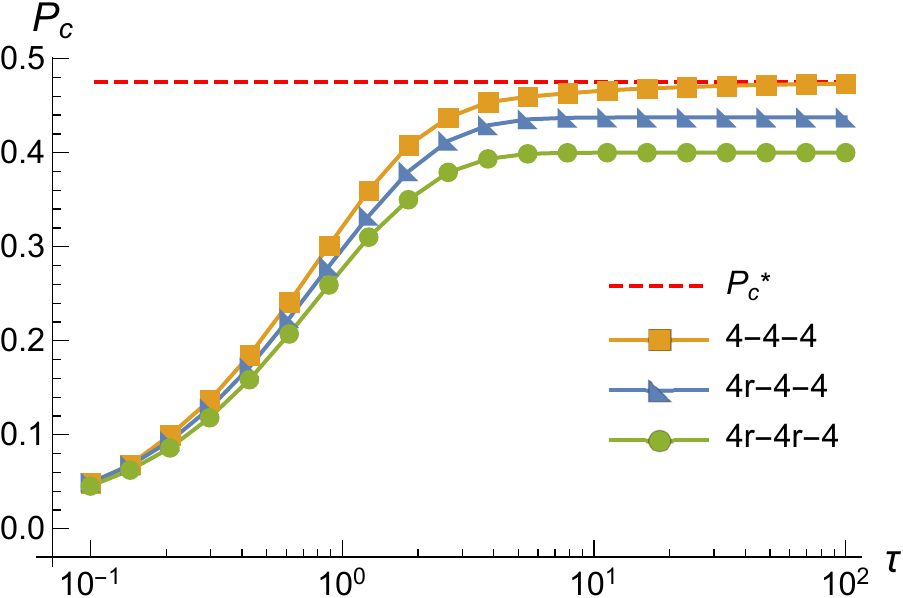}
}%
\\[2mm]
\subfloat[$\alpha = 0.7$  \label{fig:4-_07}]{%
	\centering
	\includegraphics[width=0.9\linewidth]{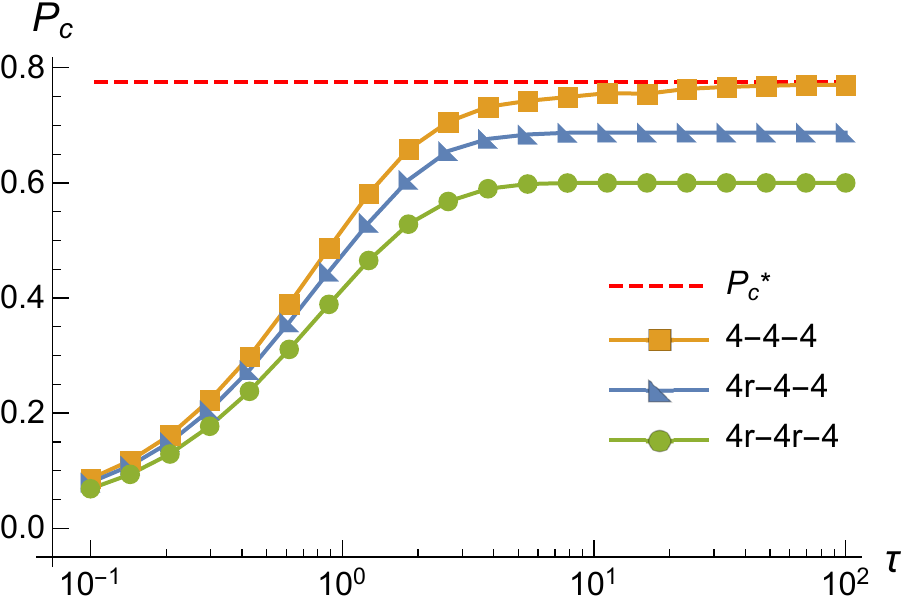}
}
\caption{Probability of correct detection $P_c$ for $\mathcal{M}=4$ as a  function of time $\tau$ for different variants from a $4-4-4$ model. The quantum states to discriminate are a mixture of a pure state and the completely mixed state with a factor $\alpha, \ 1-\alpha$ as in Eq. \eqref{def:combinationStates}. %The value of $\alpha$ employed is reported in the subcaption.
}
\label{fig:models_4-4-4}
\end{figure}

As a second case, we consider the discrimination with $\mathcal{M}=M$. We define the initial quantum states  as a linear combination of pure states and the completely mixed state, i.e.
\beq
\rho^{(m)} = \alpha \pure{\varphi_m} + (1-\alpha)\frac{I}{\mathcal{M}}, \quad m=1 \ldots \mathcal{M}
\label{def:combinationStates}
\eeq
with $\ket{\varphi_m}$ being the $m$-th state in the mutually unbiased basis of the input nodes, i.e.
\beq
\ket{\varphi_m} =  \frac{1}{\sqrt{\mathcal{M}}} \sum_{k=1}^\mathcal{M} \ee^{-\ii \frac{2 \pi  m k}{\mathcal{M}}} \ket{k}
\label{def:mutualBasis}
\eeq
where $\ket{k}$ is the quantum state associated with the $k$-th input node. For $\alpha=1$ it leads to discrimination of the pure orthogonal states $\ket{k}$. 
%\ndp{numericamente si ottiene Pc=1 solo per $\mathcal{M} = 4$ e non per 8 - unknown reason} 
On the other hand, $\alpha=0$ means that $\rho^{(m)} = I/\mathcal{M}$ for all $m$, resulting in a completely random estimation with ${P_c}^*=1/\mathcal{M}$. With an intermediate value of $\alpha$ we want to simulate a noisy preparation of the states \eqref{def:mutualBasis}. We consider $\alpha=0.3$ and $\alpha=0.7$, and we show the performance for different variants of the model $4-4-4$ in Fig. \ref{fig:models_4-4-4}. The behaviour of the probability of correct detection as a function of $p$ and $\tau$ is similar to the binary case. In the case of the model $4-4-4$ we approach asymptotically the optimal theoretical bound for $p=0$, while for the models $4r-4-4$ and $4r-4r-4$ there is a gap which is more or less emphasized depending on the value of $\alpha$. 

%\begin{figure}[h!]
%\subcaptionbox{Model $4r-4r-4$, $\alpha = 0.3$ \label{fig:Pc_4r-4r-4_03_mesh}}{%
	%\centering
	%\includegraphics[width=.45\linewidth]{Pc_4-4-4_03_mesh}
%}%
%\hfill
%\subcaptionbox{Model $4r-4r-4$, $\alpha = 0.7$  \label{fig:Pc_4r-4r-4_07_mesh}}{%
	%\centering
	%\includegraphics[width=.45\linewidth]{Pc_4-4-4_07_mesh}
%}%
%\\[2mm]
%\subcaptionbox{Model $4r-4-4$, $\alpha = 0.3$  \label{fig:Pc_4r-4-4_03_mesh}}{%
	%\centering
	%\includegraphics[width=.45\linewidth]{Pc_4-4f-4_03_mesh}
%}%
%\hfill
%\subcaptionbox{Model $4r-4-4$, $\alpha = 0.7$  \label{fig:Pc_4r-4-4_07_mesh}}{%
	%\centering
	%\includegraphics[width=.45\linewidth]{Pc_4-4f-4_07_mesh}
%}%
%\\[2mm]
%\subcaptionbox{Model $4-4-4$, $\alpha = 0.3$  \label{fig:Pc_4-4-4_03_mesh}}{%
	%\centering
	%\includegraphics[width=.45\linewidth]{Pc_4f-4f-4_03_mesh}
%}%
%\hfill
%\subcaptionbox{Model $4-4-4$, $\alpha = 0.7$  \label{fig:Pc_4-4-4_07_mesh}}{%
	%\centering
	%\includegraphics[width=.45\linewidth]{Pc_4f-4f-4_07_mesh}
%}%
%\caption{Probability of correct detection for $\mathcal{M}=4$, different topologies of a $4-4-4$ model. The quantum states to discriminate are a mixture of a pure state and the completely mixed state with proportion $\alpha, \ 1-\alpha$. On the left column, $\alpha=0.3$, on the right column, $\alpha =0.7$.
%}
%\label{fig:models_4-4-4}
%\end{figure}

%\begin{figure}  
%\centering
	%\includegraphics[width=\columnwidth]{Mary4.eps}
   %\caption{M-ary discrimination with $M=4$.}
	%\label{fig:Mary4}
%\end{figure}

\section{\label{sec:conclusions}Conclusions}

In this work we have applied the formalism of quantum stochastic walks on configurable networks to the problem of quantum state discrimination, inspired by the neural network approach for deep learning of classical information as images
%In this work, we have considered an open quantum system described by a network, whose evolution is defined by a QSW on the graph. On this system, we have set up a discrimination problem between the quantum states that may be used to initialize its time evolution. 
In particular, the input nodes encode the quantum states to discriminate, while the output nodes are used to guess the right answer.
%
%input of the network is represented by a set of input nodes that defines the subspace of the quantum system where the quantum states to discriminate are prepared. 
%
%subspace of the quantused to define the initial quantum states represent the input of the network, and the remaining nodes are organized in layers resembling a neural networks. The outputs of the network are sink nodes 
%
%In the optimized network, we see that the probability of correct decision increases with the total evolution time $\tau$, and that for a fixed $\tau$ the best performances are obtained lowering $p$. An interpretation of these results could be the following: the optimization tries to realize in the network the unitary transformation that rotates the initial quantum states into the measurement basis in the optimal way. \ndp{Decoherence effects can only degrade this process and therefore result in lower performances. Almeno, al momento non sono stati osservati improvements dovuti al rumore}

We test the discrimination of binary and $\mathcal{M}$--ary set of quantum states with multiple topologies, optimizing the coefficients of the Hamiltonian and the Lindblad operators to obtain the maximum probability of correct detection. The reconfigurability of the network architecture allows to optimally discriminate numerous sets of quantum states. 
We observe that the general trend of the performances is to increase with the total evolution time $\tau$, while for a fixed $\tau$ the best performances are obtained lowering $p$.
Notice that we are not observing any beneficial noise effects probably because the considered networks are very small and not homogeneous (not equal Hamiltonian and noise terms). In many cases with a pure quantum walk ($p=0$) we can asymptotically approach the optimal theoretical performance. When this happens, the optimized dynamics realize the Naimark extension (on the whole quantum system) of the optimal POVM for the discrimination. 
In some cases there is a gap between the theoretical and the asymptotic performance for two reasons, i.e., the lack of node connectivity, which prevents the realization of the optimal POVM, or the presence of an invariant subspace trapping a portion of the quantum states to discriminate, which prevents this component to reach the output nodes.
%presence of a portion of the quantum states to discriminate in an invariant subspace that is not connected to the output nodes.

We have also analyzed the robustness of the optimized network with respect to the preparation of quantum states and the setup of the optimal coefficients of the Hamiltonian and Lindblad operators. Indeed, the architecture is  very robust with respect to noise on both stages, as shown in Figs.~\ref{fig_robustness_states} and \ref{fig_robustness_states_percentError}. This analysis is promising for an experimental realization of the protocol, where imperfections in the preparation apparatus or in the network would be mitigated by the robustness of the architecture.
 
Therefore, we believe our results may represent further steps towards
quantum implementations of machine learning protocols, for instance to solve classification problem. Further studies will address the model applicability beyond discrimination problems, with larger networks, and by experimental benchmark on photonics--based architectures and in cold atoms platforms.

\section*{Acknowledgments}
This work was financially supported by Fondazione CR Firenze (projects Q-BIOSCAN and Quantum-AI), by PATHOS EU H2020 FET-OPEN grant no. 828946, and by the University of Florence grant Q-CODYCES.
%S.G., N.D.P., and F.C. were financially supported by Fondazione CR Firenze through the project Q-BIOSCAN, PATHOS EU H2020 FET-OPEN grant no. 828946, and UNIFI grant Q-CODYCES. M.M. acknowledges funding from the EC H2020 grant 820394 (ASTERIQS).

\appendix

\section{\label{vectorization}Vectorization of the master equation}

In the following appendices we analytically evaluate and optimize the time--evolution of the quantum system. This allows us to get a better insight on the performance of the optimization, and to explain the asymptotic behaviour  in the small topologies. 

We start by recalling the master equation that describes the evolution of a stochastic quantum walk on a graph connected to some sinks,
\begin{align}
\dot \rho =  & -(1-p)\ii \left[ H,\rho \right] + p \sum_{i,j}  L_{i,j} \rho L_{i,j}^{\dagger} - \frac{1}{2} \left \{ L_{i,j}^{\dagger}L_{i,j}, \rho \right \} \notag \\
%& \quad + \sum_{\hat{k}=1}^m \Gamma \Big (  2 \ket{\hat{k}}\bra{s_{\hat{k}}}\rho \ket{s_{\hat{k}}} \bra{\hat{k}} - \{ \pure{s_{\hat{k}}}, \rho \}  \Big ). \\
& \quad + \Gamma \sum_{n=1}^\mathcal{M}   2 \ket{n}\bra{s_n}\rho \ket{s_n} \bra{n} - \{ \pure{s_n}, \rho \} \ .
\label{masterEquationAppendix}
\end{align}
Eq. \eqref{masterEquationAppendix} defines a system of linear ordinary differential equations on the entries of the density matrix $\rho_{m,n}^{(x)} = \bra{m}\rho^{(x)} \ket{n}$. This conversion can easily be seen by applying the vectorization operation (by columns) on the members of \eqref{masterEquationAppendix}, exploiting the linear algebra property \cite{MagnusNeudecker}
\beq
\vectorize{ABC} = \left( C^{T} \otimes A \right)\vectorize{B}, 
\eeq
where $A, B, C$ are matrices with appropriate size. With the substitution $L_{i,j} = \sqrt{T_{i,j}} \ket{i} \bra{j}$ we obtain
\begin{align}
\vectorize{\dot \rho} & = \Big [ -(1-p)\ii (I \otimes H- H^T \otimes I)   \label{hermitianOperatorVec} \\
& \qquad + p \sum_{i,j} T_{i,j} \ket{i}\bra{j} \otimes \ket{i}\bra{j}  \label{lindblad1}\\
& \qquad - p \sum_{i,j} T_{i,j} \frac{1}{2} \left ( I \otimes \pure{j} + \pure{j} \otimes I \right) \label{lindblad2} \\
& \qquad  + \Gamma \sum_n 2 \ket{n} \bra{s_n} \otimes \ket{n} \bra{s_n}  \\
& \qquad - I \otimes \pure{n} - \pure{n} \otimes I \ \Big ] \vectorize{\rho} \\
& = \tilde{L} \ \vectorize{\rho}
\label{ODEL}
\end{align}
where $\tilde{L}$ is the matrix that collects all the terms in the square brackets.

The items of the density matrix are collected in a (column) vector, i.e., $  \vectorize{\rho} = \left [ \rho_{1,1} \rho_{1,2} \ldots \rho_{m,n} \ldots  \right ]^T $, one can apply an invertible transformation $r  = P\ \vectorize{\rho}$ that separates real and imaginary part of the off-diagonal entries, $\rho_{m,n} = a_{m,n} + \ii b_{m,n}$ with $\ m \neq n$, also rearranging the order of the items by putting the diagonal term first. 
%The construction of $P$ is detailed in Appendix \ref{appendix:P}, which also gives a numerical procedure to build it. 
This decomposition allows to rewrite Eq. \eqref{ODEL} in terms of $r$ as
\beq
\dot r = \left(P \tilde{L} P^{-1}\right) \left( P\ \vectorize{\rho} \right) = \mathcal{L} \ r.
\eeq

\section{\label{invariant}Invariant subspaces}

In what follows we assume a $2r-2r-2$ model, and we keep track of only the entries of $\rho$ corresponding to input and intermediate nodes since the population on the sinks can be evaluated from Eq. \eqref{def:SinkPopulation}. We define the Hamiltonian as
\beq
H = \left(
\begin{array}{cccc}
 0 & 0 & h_1 & h_2 \\
 0 & 0 & h_3 & h_4 \\
 h_1 & h_3 & 0 & 0 \\
 h_2 & h_4 & 0 & 0 \\
\end{array}
\right), \ h_k \in \mathbb{R}
\eeq
and the transition matrix as
\beq
T = \left(
\begin{array}{cccc}
 0 & 0 & t_1 & t_2 \\
 0 & 0 & 1-t_1 & 1-t_2 \\
 t_3 & t_4 & 0 & 0 \\
 1-t_3 & 1-t_4 & 0 & 0 \\
\end{array}
\right), 0 \leq t_k \leq 1 \ .
\label{def:HT}
\eeq

Rearranging the items of $\rho$ into $r$ (see Appendix \ref{vectorization}) shows us that the matrix $\mathcal{L}$ is block diagonal. This is due to the fact that we have assumed the coefficients of $H$ and $T$ to be real numbers. This allows us to separate the system of differential equations into two sub-systems that evolve independently, one involving the variables $\{\rho_{1,1},$ $\rho_{2,2},$ $\rho_{3,3}, \rho_{4,4},$ $a_{1,2},$ $a_{3,4},$ $b_{1,3},$ $b_{1,4},$ $b_{2,3},$ $b_{2,4}\}$, and the other one involving the variables $\{a_{1,3},$ $a_{1,4},$ $a_{2,3},$ $a_{2,4},$ $b_{1,2},$ $b_{3,4}\}$ (see Appendix \ref{vectorization} for their definitions), which are
\allowdisplaybreaks
\begin{align}
\dot \rho_{1,1} & =  - 2 h_1 (1-p) b_{1,3} - 2 h_2 (1-p) b_{1,4} \notag \\
& \quad +p \left(t_1 \rho_{3,3}+t_2 \rho_{4,4}\right) - p \rho_{1,1} \notag \\
\dot \rho_{2,2} & = - 2 h_3 (1-p) b_{2,3} - 2 h_4 (1-p) b_{2,4} \notag \\
& \quad +p \left[ \left(1-t_1\right) \rho_{3,3}+\left(1-t_2\right) \rho_{4,4}\right] - p \rho_{2,2} \notag \\
\dot \rho_{3,3} & = 2 h_1 (1-p) b_{1,3}+2 h_3 (1-p) b_{2,3} \notag \\
& \quad + p \left( t_3 \rho_{1,1}+t_4 \rho_{2,2}\right)-(2+p)\rho_{3,3} \notag \\
\dot \rho_{4,4} & = 2 h_2 (1-p) b_{1,4}+2 h_4 (1-p) b_{2,4} \notag \\
& \quad +p \left[(1-t_3) \rho_{1,1}+(1-t_4) \rho_{2,2}\right]-(p+2)\rho_{4,4} \notag \\
\dot a_{1,2} & = -(1-p) \left(h_3 b_{1,3}+h_4 b_{1,4}+h_1 b_{2,3}+h_2 b_{2,4}\right)\notag \\
& \quad -p a_{1,2} \notag \\
\dot b_{1,3} & = (1-p) \left(h_3 a_{1,2}-h_2 a_{3,4}-h_1 \left(\rho_{3,3}-\rho_{1,1}\right) \right)\notag \\
& \quad -(p+1) b_{1,3} \notag \\
\dot b_{1,4} & = (1-p) \left(h_4 a_{1,2}-h_1 a_{3,4}-h_2 \left(\rho_{4,4}-\rho_{1,1}\right)\right)\notag \\
& \quad -(p+1) b_{1,4} \notag \\
\dot b_{2,3} & = (1-p) \left(h_1 a_{1,2}-h_4 a_{3,4}+h_3 \left(\rho_{2,2}-\rho_{3,3}\right)\right)\notag \\
& \quad -(p+1) b_{2,3} \notag  \notag \\
\dot b_{2,4} & = (1-p) \left(h_2 a_{1,2}-h_3 a_{3,4}+h_4 \left(\rho_{2,2}-\rho_{4,4}\right)\right)\notag \\
& \quad -(p+1) b_{2,4} \notag \\
\dot a_{3,4} & = (1-p) \left(h_2 b_{1,3}+h_1 b_{1,4}+h_4 b_{2,3}+h_3 b_{2,4}\right) \notag \\
& \quad -(p+2) a_{3,4} \label{def:ode1}
\end{align}
and
\begin{align}
\dot a_{1,3}&= -(1-p) \left(h_3 b_{1,2}+h_2 b_{3,4}\right)-(p+1) a_{1,3} \notag\\
\dot a_{1,4}&= -(1-p) \left(h_4 b_{1,2}-h_1 b_{3,4}\right)-(p+1) a_{1,4} \notag \\
\dot a_{2,3}&= (1-p) \left(h_1 b_{1,2}-h_4 b_{3,4}\right)-(p+1) a_{2,3} \notag \\
\dot a_{2,4}&= (1-p) \left(h_2 b_{1,2}+h_3 b_{3,4}\right)-(p+1) a_{2,4} \notag \\
\dot b_{1,2}&= (1-p)(h_3  a_{1,3}+h_4a_{1,4}-h_1 a_{2,3}-h_2 a_{2,4}) \notag \\
& \quad -p b_{1,2} \notag \\
\dot b_{3,4}&= (1-p) \left(h_2 a_{1,3}-h_1 a_{1,4}+h_4 a_{2,3}-h_3 a_{2,4}\right) \notag \\
& \quad -(p+2) b_{3,4}  \label{def:ode2} 
\end{align}

In the first sub-system of differential equations the sinker nodes appear, but none of them are present in the sub-second system. This means that even if both sub-systems may have a not-null initial value in the variables
$\rho_{1,1},\ \rho_{2,2},\ a_{1,2},\ b_{1,2}$, only the components $\rho_{1,1},\ \rho_{2,2},\ a_{1,2}$ may end up in the sink. 
The component $b_{1,2}$ of the initial state, which corresponds to the $r_y$ coordinates, will not contribute to the sink population, regardless of the entries in $H,\ T,\ p$. Potential differences in this component between the initial quantum states, which could potentially help the discrimination, will not be visible at the sinks, effectively reducing the probability of correct decision. The value of $b_{1,2}$ of the initial quantum state is hence irrelevant to the discrimination performed on the sink nodes, and the problem is equivalent to discriminate the quantum states with this entry set to zero. In the literature, this phenomenon is explained in terms of invariant subspaces \cite{Caruso2009, Chin2010}, that is, a subspace that prevents the dynamics to escape from the network. When this invariant subspace does not contain any sink and it is initialized by the quantum states, it reduces the probability of correct detection since its time--evolution
is irrelevant for the discrimination.

Note that the presence of the invariant subspace is due to the topology of the  $2r-2r-2$ model. The model $2-2-2$ instead shows a greater connectivity between the nodes, and does not exhibit the separation of the ordinary differential equation system into two sub-systems that generate the invariant subspace.

\section{\label{sec:solutionP0} Solution of the master equation for $p=0$}

In this Appendix we solve the discrimination problem with $p=0$ in the case of equal probable pure states. Since $p=0$, there are no Lindblad operators in the master equation except the sink terms, i.e. vanishing terms \eqref{lindblad1} and \eqref{lindblad2}, and the solution can be obtained by finding the optimal value of $h_1,\ h_2,\ h_3,\ h_4$.

Since the term $b_{1,2}$ does not contribute to the discrimination, we will ignore the system of differential equations \eqref{def:ode2} of the invariant subspace and focus on the other system of differential equations, i.e. \eqref{def:ode1}. We can then restrict our attention to the discrimination of pure states such as
\beq
\ket{\psi^{(1)}} = 
\begin{pmatrix}
\cos(\alpha)\\
\sin(\alpha)
\end{pmatrix}, \
\ket{\psi^{(2)}} = 
\begin{pmatrix}
\cos(\beta)\\
\sin(\beta)
\end{pmatrix}, \
\alpha \neq \beta \in \left[0,\frac{\pi}{2} \right]
\eeq
which have no $r_y = \san{\psi^{(x)}}{\sigma_y}{\psi^{(x)}}$ coordinates.

It is convenient to apply a rotation to the quantum states in order to highlight the symmetry of the problem. By applying the unitary
\beq
U = 
\begin{pmatrix}
 \cos \left(\frac{\alpha+\beta}{2}\right) & \sin \left(\frac{\alpha+\beta}{2}\right) \\
 -\sin \left(\frac{\alpha+\beta}{2}\right) & \cos \left(\frac{\alpha+\beta}{2}\right) \\
\end{pmatrix},
\label{rotationU}
\eeq
we obtain 
\begin{align}
\ket{\phi^{(1)}} & = U \ket{\psi^{(1)}} = 
\begin{pmatrix}
\cos(\theta)\\
\sin(\theta)
\end{pmatrix}, \\
\ket{\phi^{(2)}} & = U \ket{\psi^{(2)}} =
\begin{pmatrix}
\cos(\theta)\\
-\sin(\theta)
\end{pmatrix}
\end{align}
with $\theta=(\alpha-\beta)/2$.

We can then proceed to solve the discrimination problem on $\ket{\phi^{(1)}},\ \ket{\phi^{(2)}}$. From its solution $H^*(\theta)$ we can recover the solution of the original problem as $H^*(\alpha,\ \beta) = U^\dagger H^*(\theta) U$. 

We also assume that the optimal solution $H^*$ verifies $h_1=h,\ h_2=h,\ h_3=h,\ h_4 = -h$, i.e.
\beq
H^{*(p=0)}=\begin{pmatrix}
0 & 0 & h & h \\
0 & 0 & h & -h \\
h & h & 0 & 0 \\
h & -h & 0 & 0
\end{pmatrix}
\label{optimalH}
\eeq
We will shortly see that under this ansatz we can optimize $h$ such that for $\tau$ going to infinity we reach the Helstrom bound.

Under these assumptions, the system of differential equation \eqref{def:ode1} separates into two subsystems with disjoint variables. For instance, $\rho_{3,3},\ s=b_{1,3}+b_{2,3},\ p=\rho_{1,1}+\rho_{2,2}+2 a_{1,2}$ form the following system of differential equations
\beq
\begin{array}{ccl}
\dot \rho_{3,3} & = & - 2\ \rho_{3,3} + 2 h\ s \\
\dot s & = & -2h\ \rho_{3,3} -s + h\ p\\
\dot p & = & -4 h\ s
\end{array}
\eeq
The same system holds for the variables $\rho_{4,4},\ d = b_{1,4}-b_{2,4} ,\ m= \rho_{1,1}+\rho_{2,2}-2 a_{1,2}$ in place of $\rho_{3,3},\ s,\ p$ respectively. In matrix form, 
\beq
\dot w = 
\begin{bmatrix} 
-2 & 2h & 0\\
-2h & -1 & h \\
0 & -4h & 0 
\end{bmatrix} \ w
\label{oderho33DM}
\eeq 
with $w=[\rho_{3,3}\ s\ p]^T$ or $w=[\rho_{4,4}\ d\ m]^T$.

A fundamental set of solutions for the system \eqref{oderho33DM} can be arranged in a matrix as
\beq
W(t) = \ee^{-t}\left(
\begin{array}{ccc}
 2 h & \left(1-4 h^2+z\right) \ee^{-z t} & \left(1-4 h^2-z\right) \ee^{z t} \\
 1 & 2 h \left(1+z\right) \ee^{-z t} & 2 h \left(1-z\right) \ee^{z t} \\
 4 h & 8 h^2 \ee^{-z t} & 8 h^2 \ee^{z t} \\
\end{array}
\right)
\label{wroskian}
\eeq
with $z=\sqrt{1-8 h^2}$. The Wronskian reads $\textrm{Det}[W(t)] =  -16 h^2 (1 - 8 h^2)^{3/2} \ee^{-3 t}$.

The initial conditions for \eqref{oderho33DM} are 
\beq
\begin{array}{ccl}
\rho_{3,3}(0) & = & 0 \\
s(0) & = & b_{1,3}(0)+b_{2,3}(0) = 0  \\
p(0) & = & \rho_{1,1}(0)+\rho_{2,2}(0)+2 a_{1,2}(0) \\
& = & 1+(-1)^x\sin(2\theta), 
\end{array}
\label{initialConditionRho33DM}
\eeq
with $x=0,1$ denoting the initial quantum state. Similarly, we have
\beq
\begin{array}{ccl}
\rho_{4,4}(0) & = & 0  \\
d(0) & = & b_{1,4}(0)-b_{2,4}(0) = 0  \\
m(0) & = & \rho_{1,1}(0)+\rho_{2,2}(0)-2 a_{1,2}(0) \\
& = & 1-(-1)^x\sin(2\theta), 
\end{array}
\label{initialConditionRho44SP}
\eeq
In particular, defining $\rho_{n,n}^{(x)} (t)= \trace{\Pi_n \rho^{(x)} (t)}$, $\rho^{(x)}(0) = \pure{\phi^{(x)}},\ x=1,2$, we obtain 
\begin{align}
\rho_{3,3}^{(x)}(t) & = \frac{4 h^2 e^{-t} \sinh ^2\left(\frac{1}{2} z t\right)}{z^2} (1+(-1)^x\sin(2\theta)),\\
\rho_{4,4}^{(x)}(t) & = \frac{4 h^2 e^{-t} \sinh ^2\left(\frac{1}{2} z t\right)}{z^2} (1-(-1)^x\sin(2\theta)).
\end{align}

\begin{figure}[h]
\centering
	\includegraphics[width=0.9\columnwidth]{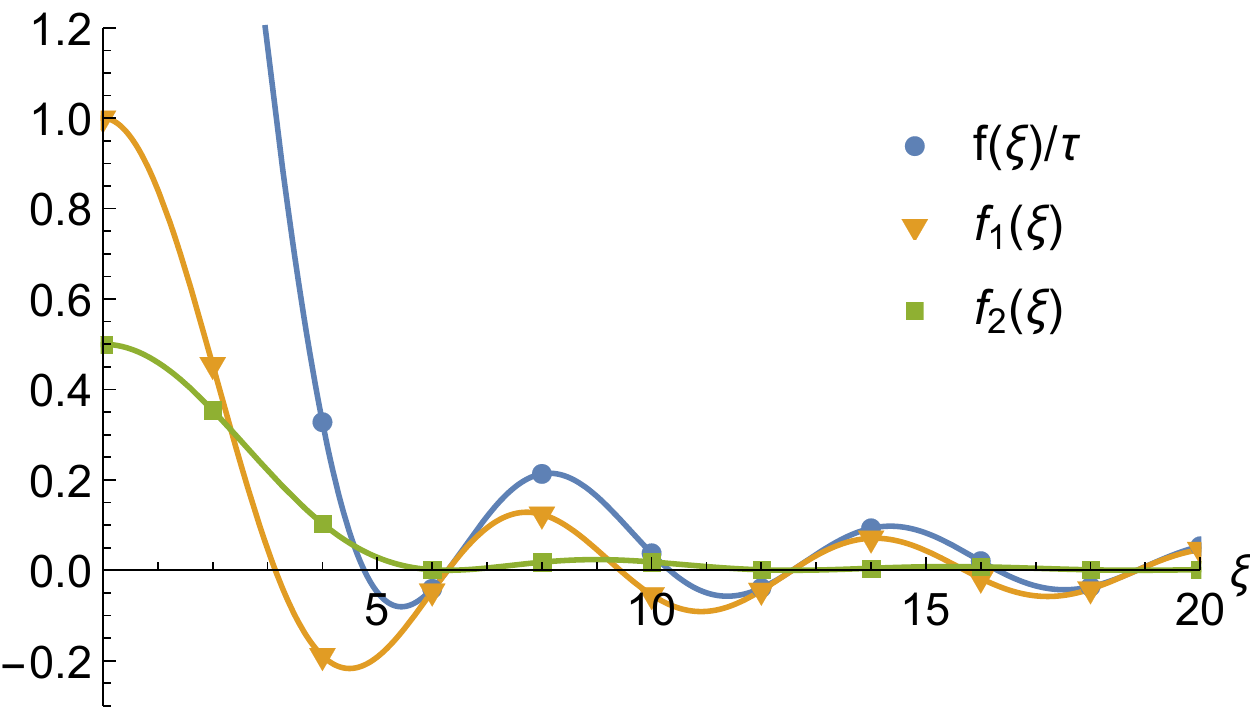}
   \caption{Plot of the term $f(\xi)/\tau$ to minimize (blue line, circle markers), along with the terms $f_1(\xi)$ (orange line, triangular markers) and $f_2(\xi)$ (green line, square markers). Both $f_1(\xi)$ and $f_2(\xi)$ are oscillating functions with the point of minimum corresponding to the first local minimum. The global minimum of $f(\xi)$ is between these two minimum, close to the minimum of $f_1(\xi)$ for $\tau\approx 0$ or the minimum of $f_2(\xi)$ for $\tau \gg 1$. In the plot the total evolution time is $\tau=5$.}
	\label{fig:fz}
\end{figure}

To solve the discrimination problem, we need to maximize the probability of correct decision. This can be written as 
\begin{align}
& {P_c}^{(p=0)}(\tau) = \int_0^\tau \rho_{3,3}^{(1)}(t) + \rho_{4,4}^{(2)}(t)\ \dd t \notag \\
& =  \frac{1+\sin(2\theta)}{2} \left[1-e^{-\tau} \left(\frac{z \sinh \left(z \tau\right)+\cosh \left(z \tau\right)-1}{z^2}+1\right)\right]
%& P_c(T) = \frac{1}{2}\trace{\Pi_5 \rho_0(T) + \Pi_6 \rho_1(T)} \notag \\
%& \qquad \quad = \int_0^T \trace{\Pi_3 \rho_0(t) + \Pi_4 \rho_1(t)}\ \dd t \notag \\
%& \qquad \quad = \int_0^T \rho_{3,3 \vert 0}(t) + \rho_{4,4 \vert 1}(t)\ \dd t \notag \\
%& =  \frac{1+\sin(2\theta)}{2} \left[1-e^{-T} \left(\frac{z \sinh \left(z T\right)+\cosh \left(z T\right)-1}{z^2}+1\right)\right]
%%
%& = \frac{1-\sin(2\theta)}{2}  \left[1-e^{-T} \left(\frac{\left(1-z\right) e^{-z T}+\left(z+1\right) e^{z T}-2}{2 \left(z^2\right)}+1\right)\right]
\label{guessingProbabilityP0}
\end{align}
It is clear that in order to maximize $P_c^{(p=0)}(\tau)$ we need to minimize the term $f(z) = (z \sinh \left(z \tau\right)+\cosh \left(z \tau\right)-1)/z^2$, $z=\sqrt{1-8 h^2}$. Unfortunately, this transcendental real function in the complex variable $z$ cannot be minimized analytically, and we need to resort to numerical methods. Graphically, we can see that the global minimum is located in the region where $1-8h^2<0$, meaning that $z$ is a pure imaginary number. Writing $z=\ii \xi/\tau$, the function to minimize becomes $f(\xi) = \tau \left [ \frac{\sin \xi}{\xi} + \tau \frac{1-\cos \xi}{\xi^2} \right ]$. Both $f_1(\xi) = \sin(\xi)/\xi$ and $f_2(\xi) =(1-\cos(\xi))/\xi^2$ are oscillating functions with the global point of minimum corresponding to the first local minimum, and the same holds for $f(\xi)$, with the point of minimum located close to those of $f_1(\xi)$ or $f_2(\xi)$ depending on $\tau$ (see Fig.~\ref{fig:fz}).

%Particular attention must be posed on the initial conditions since the function has multiple local minimum points. From its plot it seems that the global minimum is at the lower local minimum point (see Fig. \ref{fig:fz}).

The asymptotic probability of correct decision is obtained from \eqref{guessingProbabilityP0} for $\tau \to \infty$. In this case the term in brackets vanished, and we obtain the Helstrom bound \eqref{PcHelstrom} \emph{independently} of the value $h$, as long as the optimal Hamiltonian operator verifies \eqref{optimalH}.

\section{\label{sec:solutionP1} Solution of the master equation for $p=1$}

In this Appendix we solve the discrimination problem with $p=1$. In this case, the Hamiltonian operator no more contributes to the evolution of the quantum system since the term \eqref{hermitianOperatorVec} vanishes. We obtain hence a classical random walk, which can be optimize for the entries of $T$, 
\beq
T=\begin{pmatrix}
 0 & 0 & \frac{1}{2}+d_1 & \frac{1}{2}-d_2 \\[2mm]
 0 & 0 & \frac{1}{2}-d_1 & \frac{1}{2}+d_2 \\[2mm]
 \frac{1}{2}+d_3 & \frac{1}{2}-d_4 & 0 & 0 \\[2mm]
 \frac{1}{2}-d_3 & \frac{1}{2}+d_4 & 0 & 0 
\end{pmatrix}.
\label{def:Tp1}
\eeq
On the variables $d_k$ it holds the constraints
\beq
-1/2\leq d_k \leq 1/2,\ k=1, \ldots,4.
\label{dConstraint}
\eeq

\begin{figure}[h] 
\centering
	\includegraphics[width=0.9\columnwidth]{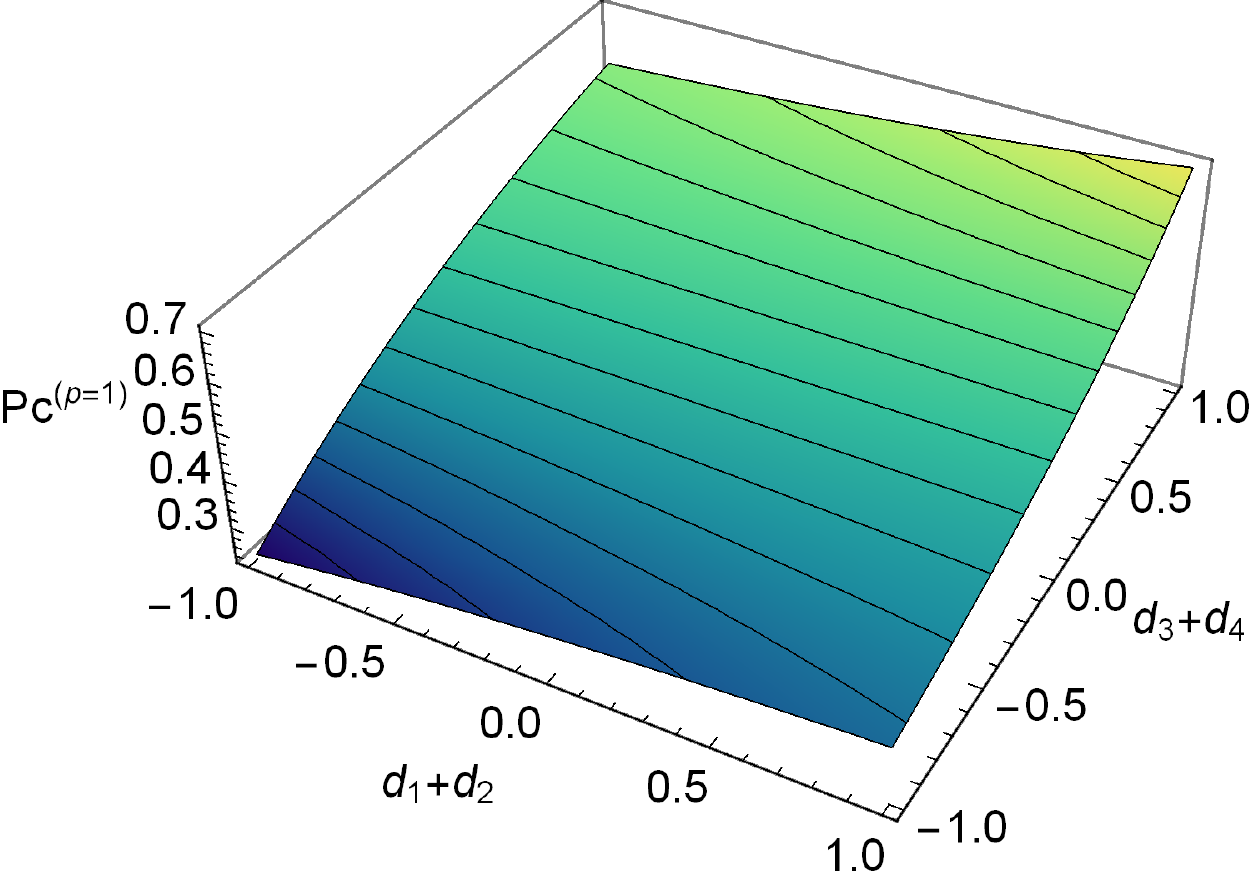}
   \caption{Probability $P_c^{(p=1)}(\tau)$ as a function of $d_1+d_2$ and $d_3+d_4$ in the case of $\Delta \rho_2-\Delta \rho_1 = 0.5,\ \tau~=~5$. The black lines are contour lines, and the maximum is obtained for  $d_1+d_2 = d_3+d_4 = \sgn(\Delta \rho_2-\Delta \rho_1)$. The same behaviour is exhibited for all $\tau$, with the surface being flatter for $\tau \approx 0$ while for $\tau \to \infty$ the surface is practically indistinguishable form the plotted one.}
	\label{fig:optimalD1d2D3d4}
\end{figure}

With $p=1$, not only the system of differential equations separates into the subsystems \eqref{def:ode1} and \eqref{def:ode2}, but also the coherence components $a_{i,j},\ b_{i,j}$ evolves independently while the diagonal terms form the coupled system
\begin{align}
\dot v & = 
\begin{bmatrix}
  -1 & 0 & \frac{1}{2}+d_1 & \frac{1}{2}-d_2 \\[2mm]
  0 & -1 & \frac{1}{2}-d_1 & \frac{1}{2}+d_2 \\[2mm]
  \frac{1}{2}+d_3 & \frac{1}{2}-d_4 & -3 & 0 \\[2mm]
 \frac{1}{2}-d_3 & \frac{1}{2}+d_4 & 0 &  -3 \\
\end{bmatrix} \ v
\label{odep1}
\end{align}
with $v=[\rho_{1,1},\ \rho_{2,2},\ \rho_{3,3},\ \rho_{4,4}]^T$.
As in the case $p=0$, we can evaluate the fundamental set of solutions and the Wronskian of the system. The initial conditions for the system \eqref{odep1} are defined from the diagonal entries of the initial states $\rho_1$ and $\rho_2$.
%\beq
%\begin{array}{ccl}
%\rho_{1,1}(0) & = &  \\
%\rho_{2,2}(0) & = &   \\
%\rho_{3,3}(0) & = & 0 \\
%\rho_{4,4}(0) & = & 0
%\end{array},
We then obtain the solutions $\rho_{3,3}^{(1)}(t),\ \rho_{4,4}^{(2)}(t)$ reported in Eqs. \eqref{r33p1}, \eqref{r44p1}, with the corresponding probability of correct decision evaluated in Eq. \eqref{guessingProbabilityP1}. As we can see, this probability depends on $\Delta \rho_2-\Delta \rho_1, \ \Delta \rho_x = \frac{\rho_{1,1}^{(1)}(0) - \rho_{2,2}^{(2)}(0)}{2}$, as well as $s_{12}=d_1+d_2$ and $s_{34}=d_3+d_4$. Notice the symmetry in $d_1, d_2\ [d_3, d_4]$. We can maximize $P_c^{(p=1)}(\tau)$ with respect to $d_1+d_2$ and $d_3+d_4$ (see Fig. \ref{fig:optimalD1d2D3d4}), and the optimal solution is $d_1+d_2 = d_3+d_4 = \sgn(\Delta \rho_2-\Delta \rho_1)$ obtained for $d_1=d_2=d_3=d_4 = \frac{1}{2}\sgn(\Delta \rho_2-\Delta \rho_1)$. For instance, in the case of $\Delta \rho_2>\Delta \rho_1$, the resulting optimal matrix $T^{*}$ reads
\beq
T^{*}=\begin{pmatrix}
0 & 0 & 1 & 0 \\
0 & 0 & 0 & 1 \\
1 & 0 & 0 & 0 \\
0 & 1 & 0 & 0
\end{pmatrix}
\label{optimalT} \ .
\eeq

Asymptotically, for $\tau \to \infty$ the optimal probability of correct decision becomes
\begin{align}
P_c^{*(p=1)}(\infty) & =\frac{1 + \vert \Delta \rho_2-\Delta \rho_1 \vert}{2} \label{PcP1Asymptotic} 
%& = \frac{1 + D[\diag(\rho_0),\diag(\rho_1)]}{2}
\end{align}
which coincides with the Helstrom bound for the discrimination of the quantum states $\rho^{(1,class.)}$ and $\rho^{(2,class.)}$, obtained from $\rho^{(1)},\ \rho^{(2)}$ by removing the coherences, effectively turning a quantum state in a classical one. 

\begin{widetext}
\begin{align}
%& \rho_{3,3 \vert x}(t) = e^{-2 t} \left \{ \frac{\sinh \left(\sqrt{2} t\right)  [1+(1-2 d_2) d_3-(1+2 d_2) d_4]}{2 \sqrt{2} [1-(d_1+d_2) (d_3+d_4)]}  \right. \label{r33p1}\\
%& \quad + \left . \frac{(d_3+d_4) \{d_1-d_2-\rho_{1,1\vert x}(0) [1-2 d_3 (d_1+d_2)]+\rho_{2,2\vert x}(0) [1-2 d_4 (d_1+d_2)] \} \sinh [ \sqrt{1+(d_1+d_2) (d_3+d_4)} \ t ]}{2 (1-(d_1+d_2) (d_3+d_4))\sqrt{1+(d_1+d_2) (d_3+d_4)}}\right \}  \notag \\
& \rho_{3,3}^{(x)}(t) = e^{-2 t} \left \{
\frac{(d_3+d_4) [d_1-d_2 + (d_1+d_2) (d_3-d_4)] \sinh \left [\sqrt{1+(d_1+d_2) (d_3+d_4)}\ t\right]}{2 [1-(d_1+d_2) (d_3+d_4)]
   \sqrt{1+(d_1+d_2) (d_3+d_4)}} \right . \label{r33p1}\\
& \qquad \left. + \frac{[1+(1-2 d_2) d_3-(1+2 d_2) d_4]\sinh \left(\sqrt{2}\ t\right)}{2 \sqrt{2} [1-(d_1+d_2)(d_3+d_4)]} 
+ \frac{(d_3+d_4) \sinh \left[\sqrt{1+(d_1+d_2)(d_3+d_4)}\ t\right]}{\sqrt{1+(d_1+d_2) (d_3+d_4)}} \Delta \rho_x
\right \} \notag \\
& \rho_{4,4}^{(x)}(t) = e^{-2 t} \left \{
-\frac{(d_3+d_4) [d_1-d_2+(d_1+d_2)(d_3-d_4)] \sinh \left[\sqrt{1+(d_1+d_2) (d_3+d_4)}\ t\right]}{2 (1-(d_1+d_2)(d_3+d_4)) \sqrt{1+(d_1+d_2) (d_3+d_4)}} \right. \label{r44p1} \\
& \qquad \left . +
\frac{[1-(1+2 d_1) d_3+(1-2 d_1) d_4] \sinh \left(\sqrt{2}\ t\right)}{2 \sqrt{2}[1-(d_1+d_2) (d_3+d_4)]}
+ \frac{(d_3+d_4) \sinh \left[ \sqrt{1+(d_1+d_2) (d_3+d_4)}\ t\right]}{\sqrt{1+(d_1+d_2)
   (d_3+d_4)}}\Delta \rho_x
\right \} \notag \\
%& \rho_{4,4 \vert x}(t) = e^{-2 t} \left \{\frac{\sinh \left(\sqrt{2} t\right) [1-(1+2 d_1) d_3+(1-2 d_1) d_4]}{2 \sqrt{2} [1-(d_1+d_2)(d_3+d_4)]}\right. \label{r44p1} \\
%& \quad \left . - \frac{(d_3+d_4) \{d_1-d_2-\rho_{1,1\vert x}(0) [1-2 d_3(d_1+d_2)]+\rho_{2,2\vert x}(0) [1-2 d_4(d_1+d_2)]\} \sinh [ \sqrt{1+(d_1+d_2)(d_3+d_4)} \ t]}{2 (1-(d_1+d_2) (d_3+d_4))\sqrt{1+(d_1+d_2) (d_3+d_4)}}\right\} \notag  \\
%
& P_c^{(p=1)}(\tau)  =  \frac{1}{2}  +	\frac{\left(\sqrt{2}-1\right) e^{-\left(2+\sqrt{2}\right) \tau}-\left(\sqrt{2}+1\right) e^{-\left(2-\sqrt{2}\right) \tau }}{4}+ \frac{(d_3+d_4) (\Delta \rho_2-\Delta \rho_1)}{3-(d_1+d_2) (d_3+d_4)} \label{guessingProbabilityP1}\\
& \qquad \qquad +\frac{(d_3+d_4) (\Delta \rho_2-\Delta \rho_1)}{\sqrt{1+(d_1+d_2)(d_3+d_4)}} \left \{ \frac{e^{  -\left[2+\sqrt{1+(d_1+d_2)(d_3+d_4)}\right]\tau}}{\sqrt{1+(d_1+d_2)(d_3+d_4)}+2}+\frac{e^{-\left[2-\sqrt{1+(d_1+d_2)(d_3+d_4)}\right]\tau}}{\sqrt{1+(d_1+d_2)(d_3+d_4)}-2}\right \}
\notag
\end{align}
\end{widetext}

\bibliography{sqwDiscrimination_noUrl}

\end{document}